\documentclass[sn-mathphys,Numbered]{sn-jnl}

\usepackage{graphicx}%
\usepackage{multirow}%
\usepackage{amsmath,amssymb,amsfonts}%
\usepackage{amsthm}%
\usepackage{mathrsfs}%
\usepackage[title]{appendix}%
\usepackage{xcolor}%
\usepackage{textcomp}%
\usepackage{manyfoot}%
\usepackage{booktabs}%
\usepackage{algorithm}%
\usepackage{algorithmicx}%
\usepackage{algpseudocode}%
\usepackage{listings}%
\usepackage{acro}
\usepackage{tikz}
\usepackage{svg}
\usepackage{placeins}
\usepackage{hyperref}
\usepackage{orcidlink}

\DeclareAcronym{rd}{
  short=RD,
  long=reaction diffusion,
}

\DeclareAcronym{af}{
  short=AF,
  long=atrial fibrillation,
}

\DeclareAcronym{fmm}{
  short=FMM,
  long=fast marching method,
}

\DeclareAcronym{fim}{
  short=FIM,
  long=fast iterative method,
}

\DeclareAcronym{at}{
  short=AT,
  long=activation time,
}

\DeclareAcronym{rt}{
  short=RT,
  long=repolarization time,
}

\DeclareAcronym{vm}{
  short=Vm,
  long=Transmembrane voltage,
}
\DeclareAcronym{erp}{
  short=ERP,
  long=effective refractory period,
}

\DeclareAcronym{dream}{
  short=DREAM,
  long=diffusion reaction eikonal alternant model,
}

\DeclareAcronym{ap}{
  short=AP,
  long=action potential,
}
\DeclareAcronym{apd}{
  short=APD,
  long=action potential duration,
}
\DeclareAcronym{re}{
  short=RE,
  long=reaction eikonal,
}

\DeclareAcronym{cv}{
  short=CV,
  long=conduction velocity,
}

\DeclareAcronym{di}{
  short=DI,
  long=diastolic interval,
}
\DeclareAcronym{pcl}{
  short=PCL,
  long=pacing cycle length,
}

\DeclareAcronym{cycfim}{
  short=cycFIM,
  long=cyclical FIM,
}
\DeclareAcronym{peerp}{
  short=PEERP,
  long=pacing at the end of the effective refractory period,
}
\DeclareAcronym{vita}{
  short=VITA,
  long=virtual induction and treatment of arrhythmias,
}
\DeclareAcronym{rtt}{
  short=RTT,
  long=round-trip time,
}

\DeclareAcronym{rmse}{
  short=RMSE,
  long=root mean square error,
}

\algdef{SE}[WHILE]{While}{EndWhile}[1]{\algorithmicwhile\ #1\ }{\algorithmicend\ \algorithmicwhile}%
\algdef{SE}[FOR]{For}{EndFor}[1]{\algorithmicfor\ #1\ }{\algorithmicend\ \algorithmicfor}%
\algdef{SE}[IF]{If}{EndIf}[1]{\algorithmicif\ #1\ }{\algorithmicend\ \algorithmicif}%

\lstdefinestyle{customalg}{
    belowcaptionskip=1\baselineskip,
    breaklines=true,
    frame=L,
    xleftmargin=\parindent,
    language=Python,
    showstringspaces=false,
    basicstyle=\footnotesize\ttfamily,
    keywordstyle=\bfseries\color{green!40!black},
    commentstyle=\itshape\color{purple!40!black},
    identifierstyle=\color{blue},
    stringstyle=\color{orange},
    numbers=left,
    stepnumber=1,
    numbersep=10pt,
    tabsize=2,
    captionpos=b,
    morekeywords={While, EndWhile, For, EndFor, If, EndIf},
}

\begin{document}

\title[Article Title]{A Cyclical Fast Iterative Method for Simulating Reentries in Cardiac Electrophysiology Using an Eikonal-Based Model}

\author[1]{\fnm{Cristian} \sur{Barrios Espinosa}\orcidlink{0000-0002-2407-5256}}

\author[2]{\fnm{Jorge} \sur{S\'{a}nchez} \orcidlink{0000-0002-0824-2691}}

\author[1]{\fnm{Stephanie} \sur{Appel}\orcidlink{0009-0003-5124-7505}}
\author[1,3]{\fnm{Silvia} \sur{Becker}\orcidlink{0000-0002-1135-6446}}
\author[1]{\fnm{Jonathan} \sur{Krau\ss}\orcidlink{0000-0001-8823-3906}}
\author[1]{\fnm{Patricia} \sur{Mart\'{i}nez D\'{i}az}\orcidlink{0009-0006-9762-2982}}
\author[1,4]{\fnm{Laura} \sur{Unger}\orcidlink{0000-0002-5614-1983}}
\author[1]{\fnm{Marie} \sur{Houillon}\orcidlink{0000-0002-6584-0233}}
\author*[1]{\fnm{Axel} \sur{Loewe}\orcidlink{0000-0002-2487-4744}}\email{publications@ibt.kit.edu}

\affil[1]{\orgdiv{Institute of Biomedical Engineering}, \orgname{Karlsruhe Institute of Technology (KIT)}, \orgaddress{\city{Karlsruhe},  \country{Germany}}}
\affil[2]{\orgdiv{ITACA Institute}, \orgname{Universitat Politècnica de València}, \orgaddress{ \city{Valencia},  \country{Spain}}}
\affil[3]{\orgdiv{Klinik f\"ur Kardiologie und Angiologie}, \orgname{Universitäts-Herzzentrum Freiburg - Bad Krozingen}, \orgaddress{ \city{Bad Krozingen},  \country{Germany}}}
\affil[4]{\orgdiv{Medizinische Klinik IV}, \orgname{Städtisches Klinikum Karlsruhe}, \orgaddress{ \city{Karlsruhe},  \country{Germany}}}

\abstract{\textbf{Background:} Computer models for simulating cardiac electrophysiology are valuable tools for research and clinical applications. Traditional reaction-diffusion (RD) models used for these purposes are computationally expensive. While eikonal models offer a faster alternative, they are not well-suited to study cardiac arrhythmias driven by reentrant activity. The present work extends the diffusion-reaction eikonal alternant model (DREAM), incorporating conduction velocity (CV) restitution for simulating complex cardiac arrhythmias.

\textbf{Methods:} The DREAM modifies the fast iterative method to model cyclical behavior, dynamic boundary conditions, and frequency-dependent anisotropic CV. Additionally, the model alternates with an approximated RD model, using a detailed ionic model for the reaction term and a triple-Gaussian to approximate the diffusion term. The DREAM and monodomain models were compared, simulating reentries in 2D manifolds with different resolutions.

 \textbf{Results:} The DREAM produced similar results across all resolutions, while experiments with the monodomain model failed at lower resolutions. CV restitution curves obtained using the DREAM closely approximated those produced by the monodomain simulations. Reentry in 2D slabs yielded similar results in vulnerable window and mean reentry duration for low CV in both models. In the left atrium, most inducing points identified by the DREAM were also present in the high-resolution monodomain model. DREAM's reentry simulations on meshes with an average edge length of 1600\,µm were 40x faster than monodomain simulations at 200\,µm.

\textbf{Conclusion:} This work establishes the mathematical foundation for using the accelerated DREAM simulation method for cardiac electrophysiology. Cardiac research applications are enabled by a publicly available implementation in the openCARP simulator.}




\maketitle

\section{Introduction}\label{sec:intro}

Computer models have provided meaningful contributions to better understand the mechanisms of cardiac arrhythmia~\cite{grandi2019computational,trayanova2023computational}. An emerging application of cardiac modeling are tissue-level simulations to guide treatments such as ablation procedures~\cite{jacquemet2016lessons}. Tissue-level simulations can be performed using \ac{rd} or eikonal models~\cite{Nagel-2022-Comparisonofpropag,pernod2011multi}. \ac{rd} models accurately capture the complex interplay between ion channels, cellular, and tissue-level behavior~\cite{Nagel-2022-Comparisonofpropag,bishop2011bidomain}. However, they often require significant computing time, even on high performance computing systems~\cite{trayanova2023computational}. Therefore, these models are hardly compatible with clinical time frames for intraprocedural decision support. Eikonal models are an alternative to investigate cardiac arrhythmias~\cite{pernod2011multi,Loewe-2019-ID12386,jacquemet2012eikonal,wallman2012comparative}. They can be 3 orders of magnitude  faster than \ac{rd} models and thus better suited for use in clinical settings (e.g. planning ablation procedures) or uncertainty quantification ~\cite{neic2017efficient,niederer2019computational}. 
However, various eikonal-based models encounter distinct challenges that impede their ability to accurately simulate cardiac arrhythmias. These challenges include the absence of repolarization and reactivation phenomena, inadequate representation of \ac{cv} and \ac{apd} restitution (i.e., their dependence on activation frequency), and the utilization of numerical methods unsuitable for anisotropic propagation~\cite{pernod2011multi,neic2017efficient,corrado2018conduction}. While not all eikonal-based models face all these limitations, a model capable of addressing these challenges simultaneously could provide a more suitable tool for studying arrhythmia compatible with clinical time frames.

Numerical solutions for the eikonal equation are computationally inexpensive due to its simple formulation and the low mesh resolution requirements. The simplest formulations of the eikonal model are only capable of simulating one activation per node. Consequently, these versions do not account for recovery or reactivation of the cardiomyocytes~\cite{neic2017efficient,pullan2002finite}. This shortcoming hinders the simulation of reentrant activity, which is a major limitation in the context of simulating arrhythmia. To overcome this problem, Pernod et al. modified the \ac{fmm} to allow reactivation of the nodes while solving the eikonal equation~\cite{pernod2011multi}. Later, Gassa et al. extended the method to enable the simulation of rotors~\cite{gassa2021spiral}. Nonetheless, when simulating anisotropic propagation in cardiac tissue using single pass methods like the regular \ac{fmm}, numercial errors can arise because the gradient directions of the eikonal equation solution do not align with the characteristic directions (i.e., the optimal trajectories). This discrepancy affects the accurate depiction of anisotropic wave propagation dynamics, particularly in regions where these directions do not lie within the same simplex in the mesh discretization. Further limitations of single pass methods like \ac{fmm} in anisotropic media are addressed in more detail by Sethian and Vladimirsky~\cite{sethian2003ordered}. Alternative methods have been proposed to solve the anisotropic eikonal equation such as the \ac{fim}, the buffered \ac{fmm}, and the anisotropic \ac{fmm} ~\cite{fu2013fast, cristiani2009fast,pernod2011multi}. 

Neic et al. used the \ac{fim} to develop the \ac{re} model, which incorporates repolarization by linking the eikonal equation with a detailed ionic model~\cite{neic2017efficient}. However, it lacks the capability to simulate reactivation and reentry. Later, Campos et al. employed the \ac{re} model in \ac{vita}, a method to investigate ventricular tachycardia~\cite{campos2022automated}. While \ac{vita} can identify areas in the heart susceptible to isthmus-dependent reentry, it can only simulate the first reentry cycle and disregards \ac{cv} restitution and functional reentry.

Iterative methods to solve the anisotropic eikonal equation are challenging when simulating reactivation and reentry phenomena. There are 2 main factors hinder the accurate simulation of reentries using iterative solution methods. First, \acp{at} can undergo multiple changes while solving the eikonal equation iteratively unlike in single pass methods. Second, \acp{at} and \acp{rt}, i.e. the time after which a node can be reactivated, have a mutual dependency. The \ac{rt} of every node in a given activation cycle depends on the \ac{erp} and the \ac{at} in the same activation cycle. Similarly, the \ac{at} depends on whether a node has fully recovered from the previous cycle's activation before the next activation attempt. Managing these conditions becomes intricate when computing the \ac{at} for the next activation cycle while constantly updating the \ac{at} from the previous activation cycle. Another important factor to reproduce physiological reentries is to incorporate  \ac{cv} restitution. This phenomenon adds additional complexity as \ac{cv} becomes dependent on the previous \ac{at}, rather than remaining constant. Therefore, a way to calculate the frequency-dependent \ac{cv} in the eikonal model must be included in the method. This can be achieved by taking the \ac{di} of previous activation cycles in the node that is being activated, or by considering the \ac{di} of the neighboring nodes~\cite{corrado2018conduction}.

This study builds on an initial version of the \ac{dream} enabling reactivation in anisotropic media through the solution of the eikonal equation using the \ac{fim}~\cite{espinosa2022diffusion}. The \ac{dream} introduced a new strategy that alternates between the eikonal and \ac{rd} models. In the proposed update of the \ac{dream}, a novel approach to \ac{cv} restitution ensures a coherent set of \ac{cv}, \ac{di}, and \ac{at} values for each revised node during the current activation cycle. This work extends the \ac{dream} by \ac{cv} restitution properties while preserving the model's other advantages.

\section{Propagation Models}\label{sec:method}

\subsection{Monodomain Model}\label{subsec:mono}

\ac{rd} models faithfully represent the propagation of the electrical wavefront through the cardiac tissue~\cite{tung1978bi,franzone2014mathematical,keener2009mathematical}. These models are the most detailed because they incorporate more physiological mechanisms than other available models. However, numerical methods used to solve the \ac{rd} equations rely on high resolution meshes, which is the main cause for their high computational cost~\cite{vigmond2008solvers,Nagel-2022-Comparisonofpropag}. The most common examples of \ac{rd} models are the bidomain and the monodomain models~\cite{potse2006comparison}. In this work, the latter is used as control to benchmark the \ac{dream}.

Derived from the bidomain equations, the monodomain model assumes equal intracellular and extracellular anisotropy ratios and is, therefore, computationally more efficient. This assumption does not hold true particularly in scenarios such as the simulation of defibrillation, where the dynamics in the extracellular space play a significant role~\cite{kandel2015electrical}. However, for the majority of cardiac electrophysiology simulations, the monodomain model proved to be adequate due to its ability to capture various electrophysiological mechanisms accurately, such as source-sink mismatch effects~\cite{Nagel-2022-Comparisonofpropag,potse2006comparison}. The equations of the bidomain model are condensed to the monodomain equation:
\begin{align}
    \beta C_{\mathrm{m}} \frac{\partial V_{\mathrm{m}}}{\partial t} &= \nabla \cdot (\boldsymbol{\sigma}_{\mathrm{m}}\nabla V_{\mathrm{m}}) - \beta (I_{\mathrm{ion}}(V_{\mathrm{m}},\overrightarrow{\eta}) - I_{\mathrm{s}}) && \text{on } \Omega \subset \mathbb{R}^3 \label{eq:monodomain} \\
    (\boldsymbol{\sigma}_{\mathrm{m}} \nabla V_{\mathrm{m}}) \cdot \overrightarrow{n}_{\mathrm{surf}} &= 0 && \text{on } \delta\Omega \label{eq:monodomain_bc}
\end{align}
where $\beta$ is the surface-to-volume ratio, $C_{\mathrm{m}}$ is the membrane capacitance, $V_{\mathrm{m}}$ is the transmembrane voltage,  $\boldsymbol{\sigma}_{\mathrm{m}}$ is the tissue conductivity tensor, $I_{\mathrm{ion}}$ is the ionic transmembrane current density, which depends on $V_{\mathrm{m}}$ and the state variables $\overrightarrow{\eta}$ that determine the behavior of the ion channels in the cell membrane and sarcoplasmic reticulum, and $I_{\mathrm{s}}$ is the transmembrane stimulus current density. $\Omega$ represents the myocardium. There is a non-flux boundary condition at $\delta\Omega$, the boundary of the domain $\Omega$. $\overrightarrow{n}_{\mathrm{surf}}$ is the outward surface normal vector.
\subsection{Eikonal Model}\label{sec:eikonal}

The anisotropic eikonal model, based on the macroscopic kinetics of wavefront propagation, seeks to determine the activation time (\ac{at}) of points within the myocardium through the following equation:~\cite{keener2009mathematical, franzone2014mathematical}:



\begin{align}
    \sqrt{\nabla T(X)^\top \mathbf{M}(X) \nabla T(X)} &= 1  && \forall X \in \Omega \label{eq:eikonal} \\
      T(X) &= T_i  && \forall X \in \Gamma_i \subset \Omega \label{eq:eikonal_bc}
\end{align}

where $T: \Omega \rightarrow \mathbb{R}_{\geq 0}\cup \{\infty\}$, maps every point in the myocardium to its corresponding \ac{at}. $\Gamma_i$ denotes the subset of points in the myocardium where the $i$-th stimulus is applied at time $T_i \in \mathbb{R}_{\geq 0}$, for $i = 1, \ldots, n_s$, with $n_s \in \mathbb{N}$ representing the total number of stimuli. Additionally, if $T_i=T_j$ then $i=j$. $\mathbf{M}: \Omega \rightarrow \mathbb{R}^{3\times3}$ maps points in the myocardium to their tensor of squared \ac{cv} defined as:
\begin{align}
   \mathbf{M}(X) &= (v_\mathrm{l}^2(X)) \overrightarrow{l} \otimes \overrightarrow{l} + (v_\mathrm{t}^2(X))\overrightarrow{t} \otimes \overrightarrow{t} + (v_\mathrm{n}^2(X))\overrightarrow{n} \otimes \overrightarrow{n}. \label{eq:ani_tensor}
\end{align}
Here, $\overrightarrow{l}$, $\overrightarrow{t}$, and $\overrightarrow{n}$ form an orthonormal system of vectors in the longitudinal, transversal, and sheet-normal directions, respectively. $\overrightarrow{l}$ is aligned with the local preferential myocyte orientation.  $v_\mathrm{l}, v_\mathrm{t},v_\mathrm{n}:\Omega \rightarrow \mathbb{R}_{\geq 0}$ assign the \ac{cv} values in their respective directions at $X$.  

The \ac{fim} is an effective approach for solving the anisotropic eikonal equation as this algorithm is particularly suited for unstructured meshes and anisotropic local \ac{cv} functions~\cite{fu2013fast}. The single-thread version of the \ac{fim} is presented in Algorithm~\ref{alg:fim}. Multi-thread versions are also available~\cite{fu2011fast,fu2013fast,jeong2008fast}. For this work, $\Omega$ is considered to be a 2D manifold embedded in $\mathbb{R}^3$. For this reason, it is assumed that $v_\mathrm{n}=v_\mathrm{t}$, however the effect of the normal component is small since most of the characteristic directions (i.e., optimal trajectories) are almost perpendicular to $\overrightarrow{n}$. To numerically approximate the viscosity solution of the eikonal equation, a triangulation $\mathcal{T} \subset \mathcal{P}(\Omega_{\mathcal{T}})$ is defined over a finite set of points $\Omega_{\mathcal{T}} \subset \Omega$, such that the convex hull of $\Omega_{\mathcal{T}}$ (i.e., the union of all triangles in $\mathcal{T}$) approximates $\Omega$. The \ac{fim} approximates the viscosity solution of the eikonal equation only at the vertices of the triangles in $\mathcal{T}$ (i.e., points in $\Omega_{\mathcal{T}}$). When mentioning a node $X \in \Omega_{\mathcal{T}}$, we refer to both a vertex in the triangulation and its position in $\mathbb{R}^3$. This notation should not cause confusion, as each simulation in this work uses only a single mesh and coordinate system.

In the first step of the \ac{fim}, the boundary conditions of the system are defined by the \ac{at} of the source nodes (Eq.~\ref{eq:eikonal_bc}). Then, the neighbors of the nodes that belong to any $\Gamma_i$ (for $i=1,\ldots, n_s$) are included in a set of active nodes $L \subset \Omega_{\mathcal{T}}$, that is initialize as $\emptyset$, which contains the list of nodes that are being updated by the local solver. This local solver is referred to as the $\mathrm{ UPDATE()}$ function. As soon as $L$ is not empty, the list iteration begins. For each iteration, every node $X$ in $L$ is updated and the previous solution for its \ac{at} is replaced. If the difference between the old and the new solution is smaller than a certain threshold $\varepsilon$, the node is then removed from $L$ and each of its neighbors, that are not currently in $L$, is analyzed. For each neighbor, a potential new solution is calculated and only replaces the old solution if the new solution is smaller (i.e., earlier) than the old solution. If this condition is fulfilled, this neighbor is added to $L$. This process is repeated until $L$ is empty.

\begin{algorithm}
    \caption{Fast Iterative Method}\label{alg:fim}
    \begin{algorithmic}
        \State $L = \emptyset$
        \For{$X \in \Omega_{\mathcal{T}}$}
          \State $T(X) \gets \infty$
          \For{$i \in {1,\ldots}, n_s $}
            \If{$X \in \Gamma_i$}
                \State $T(X) \gets T_i$
                \For{adjacent neighbor $X_{\mathrm{NB}}$ of $X$}
                    \State add $X_{\mathrm{NB}}$ to $L$
                \EndFor
            \EndIf
          \EndFor
        \EndFor                
        \While{$L \neq \emptyset$}
            \For{$X \in L$}
                \State $p \gets T(X)$
                \State $q \gets \mathrm{ UPDATE }(X)$
                \State $T(X) \gets q$
                \If{$|p-q|< \varepsilon$}
                    \For{adjacent neighbor $X_{\mathrm{NB}}$ of $X$}
                        \If{$X_{\mathrm{NB}}$ is not in $L$}
                            \State $p \gets T(X_{\mathrm{NB}})$
                            \State $q \gets \mathrm{ UPDATE } (X_{\mathrm{NB}})$
                            \If{$p>q$}
                                \State $T(X_{\mathrm{NB}}) \gets q$
                                \State add $X_{\mathrm{NB}}$ to $L$
                            \EndIf
                        \EndIf
                    \EndFor
                    \State remove $X$ from $L$
                \EndIf
            \EndFor
        \EndWhile
    \end{algorithmic}
\end{algorithm}

When solving the eikonal equation on a triangular mesh, the local solver aims to determine the smallest \ac{at} that fits the eikonal equation at a specific node $X$. For this purpose, a potential \ac{at} is calculated for every triangle containing $X$. Let $(X,Y,Z)$ be a triangle in $\mathcal{T}$, with vertices  $X,Y,Z \in \Omega_{\mathcal{T}}$: 

\begin{equation} \label{eq:lat_triangle}
    \begin{split}
        T_{Y, Z}(X) &= \min_{\lambda \in [0, 1]} \left( \lambda T(Y) + (1-\lambda) T(Z) + \frac{\sqrt{\overrightarrow{X_{\lambda}X}^\top \mathbf{D}(X)^{-1} \overrightarrow{X_{\lambda}X}}}{v_\mathrm{l}(X)} \right),\\
        \overrightarrow{X_{\lambda}X} &= \lambda \overrightarrow{YX} + (1-\lambda) \overrightarrow{ZX},\\
        \mathbf{D}(X) &= \mathbf{M}(X) \cdot v_\mathrm{l}(X)^{-2},
    \end{split}
\end{equation}

 where $T_{Y, Z}(X)$ is the potential \ac{at} that is obtained if the characteristic direction lies within the triangle $(X,Y,Z)$. Additionally, $\mathbf{D}(X)$ is a tensor that holds information about the anisotropy of conduction, and $v_\mathrm{l}(X)$ is the \ac{cv} along the longitudinal direction at node $X$. Finally, UPDATE($X$) is set as the minimum \ac{at} among all the potential \acp{at} calculated from each triangle containing $X$:

\begin{equation} \label{eq:updateX}
           \mathrm{UPDATE}(X) = \min_{(X,Y,Z) \in \mathcal{T}} T_{Y, Z}(X).
\end{equation}

\subsection{Diffusion Reaction Eikonal Alternant Model (DREAM)}\label{subsec:dream}

The \ac{dream} is a mixed model combining an approximation of the \ac{rd} model and the eikonal model. The goal of this model is to simulate reactivation patterns on meshes with lower resolutions than required for comparable simulations with \ac{rd} models, thereby increasing computational efficiency. The \ac{dream} is inspired by the \ac{re}  model~\cite{neic2017efficient} and the multi-frontal \ac{fmm}~\cite{pernod2011multi}.
 A modified version of the \ac{fim}, named \ac{cycfim}, was implemented to solve the anisotropic eikonal equation allowing reactivations by alternating with the approximated \ac{rd} model (Section~\ref{subsubsec:cycfim}).

\begin{figure}
\centering
\includegraphics[trim=60pt 250pt 60pt 80pt, clip, width=120mm]{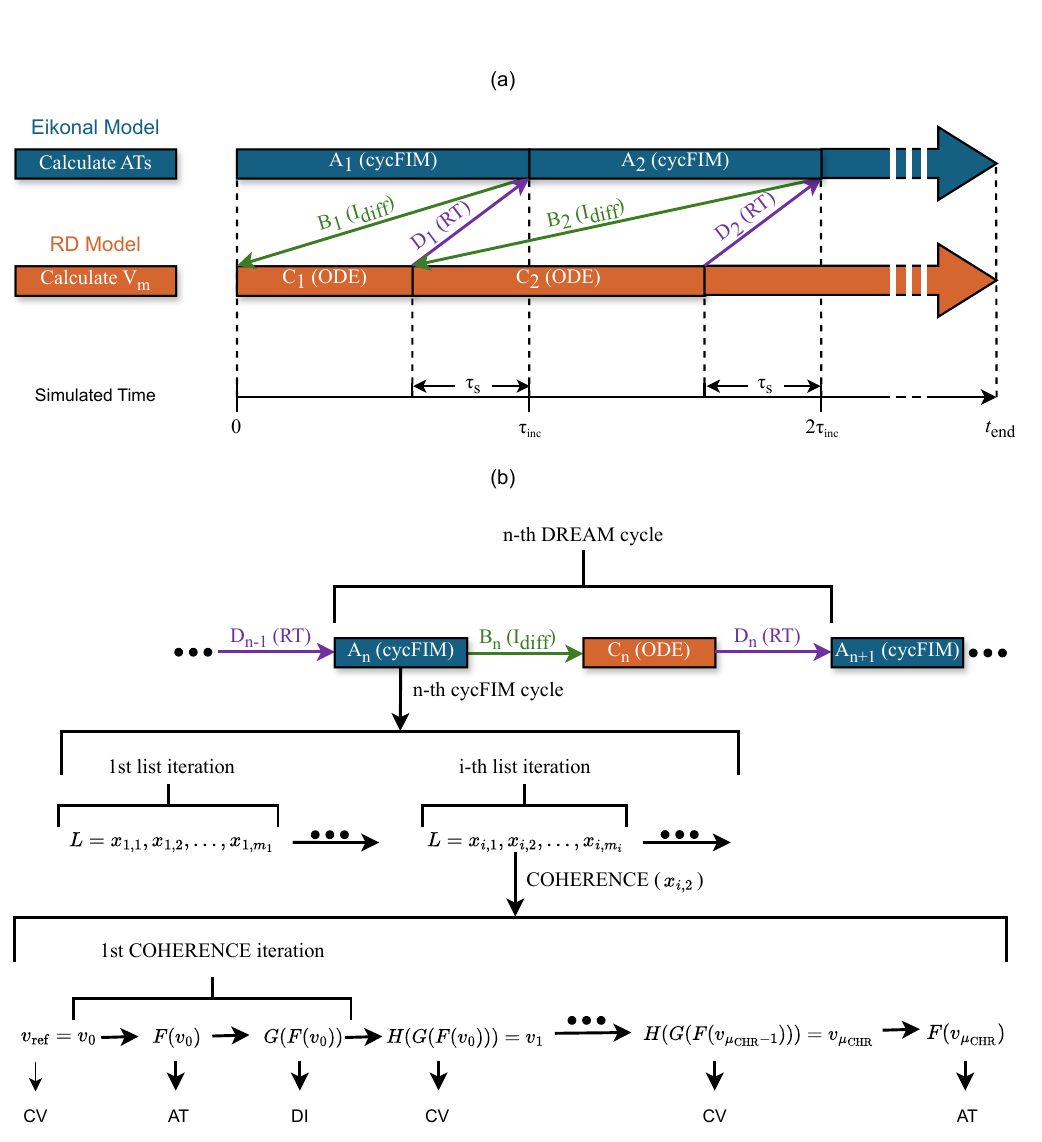}
\caption{\textbf{a)} Steps of the diffusion reaction eikonal alternant model (\ac{dream}) algorithm in simulated time: eikonal model and approximated reaction diffusion (\ac{rd}) model alternate to calculate activation times (\acp{at}) and transmembrane voltages ($V_{\mathrm{m}}$), respectively. The cyclical fast iterative method (\ac{cycfim}) is used to calculate \acp{at} (steps~$\mathrm{A}_n$). These \acp{at} are utilized to compute $I_{\mathrm{diff}}$ needed in the \ac{rd} model (steps~$\mathrm{B}_n$). $V_{\mathrm{m}}$ is determined using the approximated \ac{rd} model, by solving the ODE system of the ionic model and incorporating $I_{\mathrm{diff}}$ (steps~$\mathrm{C}_n$). $V_{\mathrm{m}}$ is used to get the repolarization times (\ac{rt}) needed in the \ac{cycfim} to allow for reactivation when solving the eikonal equation (steps~$\mathrm{D}_n$). The subscript index $n$ represents the cycle number in the sequence. $\tau_{\mathrm{inc}}$ represents the increment of $t_{\mathrm{min}}$ (minimum \ac{at} of the nodes in $L$) with each \ac{cycfim} cycle. If $L$ is never empty between stimuli, $t_{\mathrm{min}}$ tends to align approximately with multiples of $\tau_{\mathrm{inc}}$. $\tau_{\mathrm{s}}$ is the temporal safety margin to avoid conflicts between \ac{dream} cycles. \textbf{b)} Steps of the \ac{dream} algorithm, showing the sequence of a \ac{dream} cycle. In step~$\mathrm{A}_n$, each call of the \ac{cycfim} is a \ac{cycfim} cycle, containing several list iterations. $x_{i,j}$ is the $j$-th node among $m_i$ nodes in $L$ at the start of the $i$-th iteration. Visiting a node calls COHERENCE(), leading to multiple COHERENCE() iterations. In each iteration, functions $F$, $G$, and $H$ calculate \ac{at}, \ac{di}, and \ac{cv}, respectively. After $\mu_{\mathrm{CHR}}$ iterations, or upon convergence, the final output $F(v_{\mu_\mathrm{CHR}})$ is used by the \ac{cycfim}.}
\label{fig:dream}
\vspace{-10pt}
\end{figure}

Fig.~\ref{fig:dream} shows a schematic representation of the steps involved in the \ac{dream}. A single \ac{dream} cycle encompasses the execution of steps A, B, C, and D where subscript indices represent the \ac{dream} cycle's number. 
It is crucial to differentiate between the \ac{dream} cycle, comprising steps A, B, C, and D as shown in Fig.~\ref{fig:dream}b), and the concept of the activation cycle in cardiac tissue. Within a \ac{dream} cycle, each call of the \ac{cycfim} is regarded as a \ac{cycfim} cycle, which computes one \ac{at} value per node for a subset of nodes. However, it is important to note that not every review of a node will result in an activation cycle for that node, as propagation failures can occur. Additionally, not more than one \ac{at} is calculated per \ac{dream} cycle per node.

At first, the \ac{cycfim} applies the first stimulus (i.e., computes the first boundary condition) and iteratively reviews every node in $L$ and updates the \ac{at} of each node by solving the eikonal equation. A list iteration refers to the process where each node in $L$ is visited once. Therefore, every cycle encompasses one or more list iterations. At the end of each list iteration, $t_{\mathrm{min}}$ is the smallest absolute \ac{at} among all nodes in $L$. A parameter $\tau_{\mathrm{inc}}$ is defined to limit the increment of $t_{\mathrm{min}}$ throughout the list iterations of a \ac{cycfim}. Once the total increment of $t_{\mathrm{min}}$ during a \ac{cycfim} cycle exceeds $\tau_{\mathrm{inc}}$, the \ac{cycfim} cycle ends (Fig.~\ref{fig:dream}a), step~$\mathrm{A}_1$, Section~\ref{subsubsec:cycfim}). 
Hereby, $\tau_{\mathrm{inc}}$ represents the increment of $t_{\mathrm{min}}$ each time \ac{cycfim} is called. Then, the determined \acp{at} are used to calculate $I_{\mathrm{diff}}$, which approximates the diffusion current expressed in the parabolic equation of the \ac{rd} model. Then, $I_{\mathrm{diff}}$ triggers an \ac{ap} (Fig.~\ref{fig:dream}a), step~$\mathrm{B}_1$ and Section~\ref{subsubsec:Idiff}), which allows to compute $V_{\mathrm{m}}$ in a low resolution mesh. The approximated \ac{rd} model uses $I_{\mathrm{diff}}$ to compute $V_{\mathrm{m}}$ in all nodes of the mesh until $t=t_{\mathrm{min}}-\tau_{\mathrm{s}}$ (Fig.~\ref{fig:dream}a), step~$\mathrm{C}_1$ and Section~\ref{subsubsec:approxRD}). $\tau_{\mathrm{s}}$ is a safety margin in time, ensuring that only cells with converged \acp{at} are stimulated by $I_{\mathrm{diff}}$. Thereby, a reliable identification of repolarization times (\acp{rt}) is enabled. \acp{rt} are defined as the time points when $V_{\mathrm{m}}$ of an activated node $X$ crosses the threshold of $-40$\,mV with a negative slope. If $V_{\mathrm{m}}$ in an activated node does not reach this threshold before $t=t_{\mathrm{min}}-\tau_{\mathrm{s}}$, the ionic model is run independently for that node to determine its \ac{rt} (Fig.~\ref{fig:dream}a), step~$\mathrm{D}_1$ and Section~\ref{subsubsec:findRT}). To start the next cycle in step~$\mathrm{A}_2$, the \ac{cycfim} iterates again until $t_{\mathrm{min}}>t_{\mathrm{min,init}}+\tau_{\mathrm{inc}}$, where $t_{\mathrm{min,init}}$ is the last calculated $t_{\mathrm{min}}$ of step~$\mathrm{A}_1$. Hereby, the previously calculated \acp{rt} are considered by \ac{cycfim} to check whether a node can be reactivated. Then steps~$\mathrm{B}_2$, $\mathrm{C}_2$, and $\mathrm{D}_2$ are performed just as steps~$\mathrm{B}_1$, $\mathrm{C}_1$, and $\mathrm{D}_1$.
This process is repeated until $t$ reaches $t_{\mathrm{end}}$, corresponding to the end of the simulation time. Table~\ref{tab:dream_parameters} shows the variables and parameters of the \ac{dream} including their meaning and values used in this work. The code to run simulations using the \ac{dream} is available in the openCARP simulator~\cite{opencarp_eikonal}. In the next sections, each of the steps is explained in more detail.

\begin{table}[h]
    \centering
    \caption{\ac{dream} parameters and variables: Absolute times denote specific time points, with values spanning over the entire duration of the simulation. Relative times, fixed and typically smaller, constrain the occurrence of absolute times within specified intervals. Maximum permitted iteration parameters set limits for iterative processes within the \acl{cycfim}.}
    \begin{tabular}{cp{8cm}c} 
        \toprule
        \toprule
        \multicolumn{3}{c}{\textbf{Absolute times}} \\
        \midrule
        \textbf{Variable} & \textbf{Meaning} & \textbf{Value} \\
        $t_{\mathrm{min}}$ & Minimum \ac{at} of nodes in $L$ at the end of each list iteration & (-)\footnotemark[1] \\
        $t_{\mathrm{min,init}}$ & Last $t_{\mathrm{min}}$ from the previous \ac{dream} cycle & (-)\footnotemark[1] \\
        $t$ & Current time step in \ac{rd} model & (-)\footnotemark[1] \\
        \midrule
        \textbf{Parameter} & \textbf{Meaning} & \textbf{Value} \\
        \midrule
        $t_{\mathrm{end}}$ & End of simulation & (-)\footnotemark[2] \\
        \multicolumn{3}{c}{} \\
        \toprule
        \toprule
        \multicolumn{3}{c}{\textbf{Relative times}} \\
        \midrule
        \textbf{Parameter} & \textbf{Meaning} & \textbf{Value}  \\
        \midrule
        $\tau_{\mathrm{inc}}$ & Maximum allowed increment of $t_{\mathrm{min}}$ in every \ac{dream} cycle & 90\,ms \\
        
        $\tau_{\mathrm{max}}$ & Maximum allowed difference of \ac{at} values among nodes in $L$ & 90\,ms \\
        
        $\tau_{\mathrm{s}}$ &  Minimum allowed difference between $t_{\mathrm{min}}$ and $t$ & 10\,ms \\
        
        $\varepsilon$ & Threshold of convergence & 0.01\,ms \\
        \multicolumn{3}{c}{} \\
        \toprule
        \toprule
        \multicolumn{3}{c}{\textbf{Maximum iterations}} \\
        \midrule
        \textbf{Parameter} & \textbf{Meaning} & \textbf{Value} \\
        \midrule
        $\mu_\mathrm{L_1}$ & Maximum number of list iterations per node per entry into $L$ & 50\\
        
        $\mu_\mathrm{L_2}$ & Maximum number of node returns to $L$ per activation & 50\\
        
        $\mu_\mathrm{CHR}$ & Maximum number of COHERENCE() iterations per node per call& 50\\
        \bottomrule

    \end{tabular}
    \footnotetext[1]{Variable values change throughout the simulation.}
    \footnotetext[2]{$t_{\mathrm{end}}$ changes across experiments.}
    \label{tab:dream_parameters}
\end{table}

\subsubsection{Cyclical Fast Iterative Method}\label{subsubsec:cycfim}
The \ac{cycfim} is called in step~$\mathrm{A}_n$ of the \ac{dream}  algorithm with $n\in \mathbb{N}$ representing the number of the cycle. Algorithm~\ref{alg:fim_mod} describes this \ac{cycfim}. Some modifications were made to the single-thread \ac{fim} to allow alternation between the eikonal and \ac{rd} modelss. Firstly, the boundary conditions for the eikonal equation need to be dynamically applied as their effect on the system varies depending on the refractory state of the nodes receiving the stimuli. Secondly, list iterations should handle \ac{rt} and manage multiple \acp{at} per node. Thirdly, the local solver must ensure coherence among the \ac{cv}, \ac{di}, and \ac{at} for each node. If $L$ is emptied before all the boundary conditions are computed, further aspects must be considered since $t_{\mathrm{min}}$ does no longer align with multiples of $\tau_{\mathrm{inc}}$(see Section~\ref{subsubsec:emptylist}).

\paragraph{Dynamic Boundary Conditions}\label{para:dyn_bc}

Dynamic boundary conditions handle stimuli applied to specific areas of the tissue. These conditions vary dynamically as they depend on the refractory state of stimulated nodes when stimuli are applied. Unlike conventional eikonal models where boundary conditions are computed a priori irrespective of their timing, the \ac{cycfim} computes boundary conditions (i.e, stimuli) progressively during the simulation. For this reason, to compute boundary conditions in the \ac{cycfim} the order in which stimuli are applied must be considered. Let $ T_1 < T_2 < \ldots < T_{n_s} $ be the times when stimuli are applied, and let $ \Gamma_1, \ldots, \Gamma_{n_s} \subset \Omega_{\mathcal{T}} $ be the respective subsets of nodes where this stimuli are applied. At the beginning of the $n$-th \ac{cycfim} cycle, let $ T_s $, for $ s \in \mathbb{N} $ and $ 1 \leq s \leq n_s $, be the time of the $s-th$ stimulus such that $ T_1, \ldots, T_{s-1} $ have already been computed during the previous \ac{cycfim} cycles. Moreover, assume that $ T_{s}, \ldots, T_{n_s} $ have not yet been computed. Therefore, $ T_s $ denotes the time of the earliest stimulus that has yet to be processed. Boundary conditions might be computed in 2 scenarios: at the beginning of each cycle if $L$ is empty, or at the beginning of each list iteration if $L$ is not empty.

 In the first case, $L$ can be empty for 2 reasons: the simulation has just started and $n=s=1$ (i.e., the first stimulus must be computed during the first \ac{cycfim} cycle), or all activation times (\acp{at}) associated with activation waves triggered by stimuli applied before $T_s$ have converged, and $L$ was emptied during the $(n-1)$-th \ac{cycfim} cycle. To incorporate new nodes into $L$ and start a new list iteration, the $s$-th stimulus must be computed. In the second case, $L$ is not empty and new stimuli might be applied. For instance, if $T_s$ falls within the range of allowed activation times (\acp{at}) for the nodes in $L$ during this \ac{cycfim} cycle (i.e., $T_s \in [t_{\mathrm{min,init}}, t_{\mathrm{min,init}} + \tau_{\mathrm{inc}} + \tau_{\mathrm{max}}]$), then the $s$-th stimulus can be computed. The parameter $\tau_{\mathrm{max}}$ which imposes an upper limit on the \acp{at} of nodes in $ L$ will be explained in greater detail in the following subsection. Additionally, all the subsequent stimuli which times $T_{s+1}, T_{s+2}, \ldots$ are within this range, can be computed together.
 
 To compute the $s$-th stimulus in both aforementioned cases, each node where the stimulus is applied is considered independently. Let $X \in \Gamma_s$ be a node where the $s$-th stimulus is applied at time $T_s$, and let $T(X)$ and $R(X)$ be its last computed \ac{at} and \ac{rt}, respectively. Then, $T_s$ is assigned to $T(X)$ if one of the following conditions is fulfilled:
\begin{equation}
    \left(T(X)<R(X)<T_s\right) \lor \left(R(X)<T_s<T(X)\right).
\end{equation}    

 In the first condition, the calculation for the current activation of $X$ has not yet begun and that is the reason why $T(X)<R(X)$. In that case, the solution $T(X)$ is replaced if $T_s>R(X)$, indicating that at $T_s$, the node has already recovered from the previous activation and is ready to be activated again. In the second condition, $R(X)<T(X)$, i.e., the calculation for the current activation for $X$ has already begun and it is still converging. The solution is updated if $R(X)<T_s$ signifying that the node $X$ is ready to be activated at $T_s$ and ${T_s<\;T(X)}$, which means $T_s$ is a better (i.e., earlier) solution. The nodes with replaced \acp{at} are removed from $L$, while their neighbors are added to $L$. To faithfully represent occurrence of unidirectional propagation, nodes that are successfully activated by a given stimulus cannot continue to activate nodes where an activation by this stimulus was unsuccessful before.

\paragraph{List Iteration}\label{para:L_iteration}
 
 $t_{\mathrm{min,init}}$ represents the minimum \ac{at} among the nodes in $L$  reached at the end of the previous \ac{dream} cycle. When $L$ is not empty after computing the first stimulus, $t_{\mathrm{min,init}}$ is initialized to $0$ during the first call of \ac{cycfim} in the first cycle, or adjusted to the value of $t_{\mathrm{min}}$ calculated at the end of step~$\mathrm{A}_{n-1}$ if $n>1$. $t_{\mathrm{min,init}}$ also serves as a reference to be compared to $t_{\mathrm{min}}$, the smallest \ac{at} in $L$ at the end of the previos \ac{cycfim} cycle. $L$ undergoes multiple iterations until $t_{\mathrm{min}}>t_{\mathrm{min,init}}+\tau_{\mathrm{inc}}$. Then, the function COHERENCE() (see next subsection) is applied to each node in $L$. Let $X$  be the node that is being revised and $T(X)$ and $R(X)$ its last computed \ac{at} and \ac{rt}, respectively. Then, a new solution $q$ is computed by COHERENCE(X) to potentially replace $T(X)$. If the difference between the old and new \ac{at} is smaller than the threshold $\varepsilon$, or either the old or the new solution is infinite, then $X$ is removed from $L$. To avoid that a node is activated twice in the same iteration of the list, an upper bound $\tau_{\mathrm{max}}$ is applied to the difference between the maximum \ac{at} of the nodes in $L$ and $t_{\mathrm{min}}$. If a certain node has a calculated \ac{at} above the maximum allowed \ac{at} for a given iteration of \ac{cycfim}, i.e., $q>t_{\mathrm{min,init}}+\tau_{\mathrm{inc}}+\tau_{\mathrm{max}}$, then this node will not be removed from $L$ and its neighbors will not be visited yet. 
  
 When a node is removed from $L$, its neighbors that are not in $L$ are revised as well. Let $X_{\mathrm{NB}}$ be a neighbor of $X$ with last computed \ac{at} and \ac{rt} being $T(X_{\mathrm{NB}})$ and $R(X_{\mathrm{NB}})$, respectively. If $X_{\mathrm{NB}}$ is not in $L$ and a new solution $q_{\mathrm{NB}}$ is computed, one of the following conditions (similar to those for stimuli) must be fulfilled for $q_{\mathrm{NB}}$ to replace $T(X_{\mathrm{NB}})$:
 \begin{equation}
    \left(T(X_{\mathrm{NB}})<R(X_{\mathrm{NB}})<q_{\mathrm{NB}}\right) \lor
    \left(R(X_{\mathrm{NB}})< q_{\mathrm{NB}}<T(X_{\mathrm{NB}})\right).
\end{equation}
 
\ In the first condition, the calculation for the current activation of $X$ has not yet begun. In that case, the solution is replaced if $q_{\mathrm{NB}}>\;R(X_{\mathrm{NB}})$, meaning that the node has recovered at $q_{\mathrm{NB}}$. In the second condition, $R({X_{\mathrm{NB}}})<\;T(X_{\mathrm{NB}})$, i.e., calculation of the current activation has already started for $X_{\mathrm{NB}}$ and is still converging. The solution is replaced if RT$_{X_{\mathrm{NB}}}<q_{\mathrm{NB}}<\;T(X_{\mathrm{NB}})$,  which means that the node is ready to be activated at $q_{\mathrm{NB}}$  and $q_{\mathrm{NB}}$  is a better (i.e., earlier) solution. If  $q_{\mathrm{NB}}$ replaces $T(X_{\mathrm{NB}})$ in any of the 2 conditions, then $X_{\mathrm{NB}}$ is added to the list of active nodes $L$. Regardless of the previous conditions, if $q_{\mathrm{NB}}$ is infinite or smaller than $t$, $X_{\mathrm{NB}}$ will not be added to $L$. When all the nodes in $L$ and the neighbors of converging nodes have been revised, $t_{\mathrm{min}}$ is recalculated to decide whether further iterations are required in step~$\mathrm{A}_n$. To reduce the computational cost and to prevent nodes from indefinitely exiting and returning to $L$ or iterating inside of $L$, 2 limits were defined: $\mu_\mathrm{L_1}$ and $\mu_\mathrm{L_2}$. $\mu_\mathrm{L_1}$  represents the maximum number of list iterations per node per entry into $L$, and $\mu_\mathrm{L_2}$, denotes the maximum number of returns to $L$ per node per activation.

\begin{algorithm}
\caption{Cyclical Fast Iterative Method within the \textit{n}-th \ac{dream} cycle}\label{alg:fim_mod}
\begin{algorithmic}
\If{$L = \emptyset$}
\State Compute boundary conditions
\EndIf
\If{$n = 0 $}
\State $t_{\mathrm{min,init}} = 0$
\Else
\State $t_{\mathrm{min,init}} = t_{\mathrm{min}}$
\EndIf
\Repeat
\State Compute boundary conditions
\For{$X \in L$}
\State $p \gets T(X)$
\State $q \gets \mathrm{ COHERENCE }(X)$
\State $T(X) \gets q$
\If{$|p-q|< \varepsilon$}
\For{adjacent neighbor $X_{\mathrm{NB}}$ of $X$}
\If{ $X_{\mathrm{NB}} \text{ is not in } L$}
\State $p \gets T(X_\mathrm{NB})$
\State $q_{\mathrm{NB}} \gets \mathrm{ COHERENCE } (X_{\mathrm{NB}})$
\If{ $( p<R(X_{\mathrm{NB}}) <q_{\mathrm{NB}} \lor  R(X_{\mathrm{NB}})<q_{\mathrm{NB}}<p) \land (t<q_{\mathrm{NB}}<\infty)$ }
\State $T(X_{\mathrm{NB}}) \gets q_{\mathrm{NB}}$
\State add $X_{\mathrm{NB}}$ to $L$
\EndIf
\EndIf
\EndFor
\State remove $X$ from $L$
\EndIf
\EndFor
\State $t_{\mathrm{min}}=\mathrm{min}\{T(X)|X\in L\}$ 
\Until{$L = \emptyset$ OR $(t_{\mathrm{min}}-t_{\mathrm{min,init}})\geq\tau_{\mathrm{inc}}$}
\end{algorithmic}
\end{algorithm}

\paragraph{Coherence Between Conduction Velocity, Diastolic Interval and Activation Time}\label{para:modlocalsolv}

The local solver for the \ac{dream} seeks to calculate the \ac{at} for each node. Unlike the standard eikonal model, the \ac{dream} accounts for the \ac{cv} restitution at each node in each activation cycle instead of a fixed \ac{cv}. To implement \ac{cv} restitution, the function UPDATE() in Algorithm~\ref{alg:fim} is replaced by a new function COHERENCE() in Algorithm~\ref{alg:fim_mod}. The \ac{cv} restitution is incorporated in the model by providing a template restitution curve, for example calculated from monodomain model simulations or inferred from clinical data~\cite{Loewe-2019-ID12386}. To integrate this \ac{cv} restitution phenomenon into the \ac{dream}, the feedback loop among \ac{at}, \ac{cv}, and \ac{di} must be acknowledged. This feedback loop means that calculating the \ac{at} depends on the \ac{cv}, which relies on the \ac{di}, which in turn is influenced by the \ac{at}. 

When applying the COHERENCE() function to a given node $X$, the reference \ac{cv} of $X$, the \ac{rt} of $X$ from the previous activation, and \acp{at} of the neighboring nodes in the current activation are known. The reference \ac{cv} corresponds to the maximum possible \ac{cv}, i.e. $v_\mathrm{l}(X)$ when the \ac{di} is sufficiently long and \ac{cv} becomes unaffected by restitution effects. Restitution curves with biphasic behavior, where the maximum \ac{cv} is reached at intermediate \acp{di} instead of long \acp{di}, are excluded from this method. If an initial \ac{at} is estimated using the highest value of the reference \ac{cv}, this \ac{at} will have the smallest (i.e., earliest) possible value, resulting in the shortest possible \ac{di}. If this \ac{di} is short, when applying the restitution curve, this leads to a slower \ac{cv} than the reference \ac{cv} and therefore in a higher (i.e., later) \ac{at} than the initially estimated \ac{at}. To calculate a potential \ac{at} for $X$, the COHERENCE() function is utilized to iteratively recalculate \ac{cv}, \ac{di}, and \ac{at}  until a stable state of coherence between these 3 variables is reached. Note that these temporary values of the 3 variables are referring to the current activation of $X$. These 3 values are recalculated in each COHERENCE() iteration (not to be confused with list iteration see Fig.~\ref{fig:dream}b). Each application of the local solver involves one or more COHERENCE() iterations, starting with the reference \ac{cv} and ending with a potential value for $T(X)$ to be used in \ac{cycfim}. Additionally, the term \ac{cv} can refer to the speed of the wavefront itself, the longitudinal \ac{cv} when the wavefront moves in the preferential cardiomyocyte orientation, or the transversal \ac{cv} when the wavefront moves perpendicularly to this orientation. Although the COHERENCE() function utilizes the longitudinal \ac{cv} (i.e., $v_{\mathrm{l}}(X)$) for its computation and equilibrium is reached among \ac{at}, \ac{di}, and longitudinal \ac{cv}, this equilibrium also extends to the wavefront's \ac{cv}.

\begin{figure}
\centering
\includegraphics[trim=100pt 230pt 70pt 220pt, clip, width=120mm]{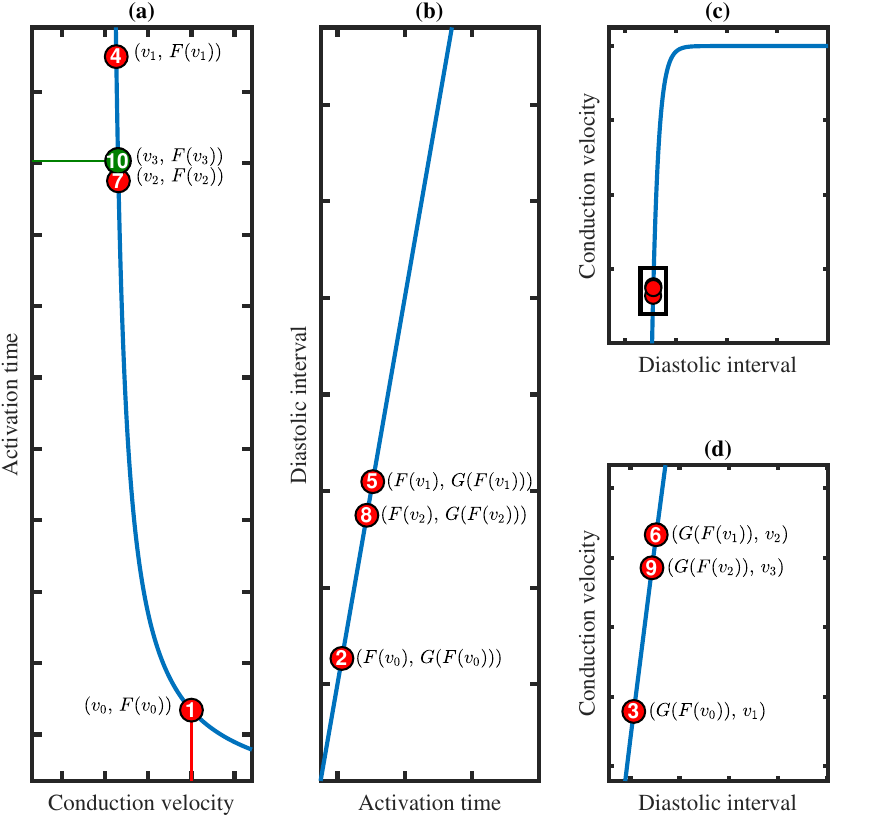}
\caption{COHERENCE() call with 3 iterations. Numbers indicate the sequence in which activation times (ATs), diastolic intervals (DIs), and conduction velocities (CVs) are calculated. Red circles correspond to temporary values and the green circle indicates the final output. \textbf{a)} Function $ F $ mapping \ac{cv} to \ac{at}. The vertical red line marks the initial input of COHERENCE() (i.e., $v_{\mathrm{ref}}=v_0$) on the horizontal axis. The green red line marks the final output of COHERENCE() (i.e, $F(v_3)$) on the vertical axis, \textbf{b)} Function $ G $ mapping \ac{at} to \ac{di}, \textbf{c)} Function $ H $ mapping \ac{di} to \ac{cv}, \textbf{d)} Zoom-in of the boxed area in \textbf{c)}, illustrating temporary \ac{di} and \ac{cv} values.}
\label{fig:coherence}
\end{figure}

COHERENCE() calculates the potential \ac{at} of $X$ using the functions $F$, $G$ and $H$. Each of these functions also depends on $X$. To simplify notation, $X$ as an argument of these functions will be omitted. Let $ F: \mathbb{R}_{\geq 0} \rightarrow \mathbb{R}_{\geq 0}\cup \{\infty\} $ be the function that computes the \ac{at} from the \ac{cv}:

\begin{equation} \label{eq:function_lat}
    \begin{split}
        F(v^*) = t^*,
    \end{split}
\end{equation}

where  $v^*$ is a temporary value for $v_\mathrm{l}(X)$ and $t^*$ is a temporary solution to the eikonal equation (Eq.~\ref{eq:eikonal}) at node $ X \in \Omega$ providing a potential \ac{at} for this node.
To solve the eikonal equation, Eq.~\ref{eq:lat_triangle} is used to collect all potential candidates for  $t^* $. $F$ is similar to the UPDATE() function of \ac{fim} but considers additional conditions. Propagation block can occur in \ac{cycfim}, implying that $v=0$, thus $F(0)=\infty$. On the other hand, when taking into account the neighboring nodes $Y$ and $Z$ for Eq.~\ref{eq:lat_triangle}, the triangle $XYZ$ is not considered if $R(Y)>T(Y)$ or $R(Z)>T(Z)$. This implies that the \acp{at} for these nodes are not yet computed for the current activation cycle and, therefore, must be excluded from the calculation of $ t^* $. Figure~\ref{fig:coherence}a shows a schematic representation of function F()'s morphology.

Now, let $ G: \mathbb{R}_{\geq 0}\cup \{\infty\} \rightarrow \mathbb{R}_{\geq 0}\cup \{\infty\} $ be the function that computes the \acs{di}  from $ t^* $:

\begin{equation} \label{eq:function_di}
    \begin{split}
        G(t^*) = d^* = t^* - R(X),
    \end{split}
\end{equation}
where $ R(X) $ is the \ac{rt} of node $ X $ from the previous activation cycle and $ d^* $ is a temporary value for \ac{di} in the current activation cycle. Figure~\ref{fig:coherence}b shows a schematic representation of function G()'s morphology.

Finally, let $ H: \mathbb{R}_{\geq 0}\cup \{\infty\} \rightarrow \mathbb{R}_{\geq 0} $ be the function that computes the new temporary \ac{cv} $ v^* $ from the temporary \ac{di} $ d^* $:

\begin{equation}\label{eq:function_cv}
    H(d^*) = v^* = \begin{cases}
         v_{\mathrm{ref}} \cdot \left(1 - \rho \cdot e^{-\frac{\log(\rho)}{\psi}(d^* + \kappa)}\right) & \text{if } d^* > \theta ,\\
         0 & \text{if } d^* \leq \theta.
    \end{cases}
\end{equation}

Here, $ v_{\mathrm{ref}} $ represents the reference \ac{cv}, while $ \rho $, $ \kappa $, $\theta$, and $ \psi $ are predefined parameters chosen to fit restitution curves for example obtained from monodomain simulations. The formulation in Eq.~\ref{eq:function_cv} allows expressing the steepness of the restitution curve in terms of $\rho$, the shift in the x-axis (i.e., \ac{di}) in $\kappa$ and the shortest propagating \ac{di} in $\theta$. Figure~\ref{fig:coherence}c shows a schematic representation of function H()'s morphology.

A recursive sequence $(v_i)$ can be defined using the composition of the functions $H$, $G$ and $F$:
\begin{equation} \label{eq:seq_convergence}
    \begin{split}
        v_0 &= v_{\mathrm{ref}},\\
        v_{n+1} &= H(G(F(v_n))).
    \end{split}
\end{equation}

The COHERENCE() function approximates the \ac{at} that corresponds to the \ac{cv} value which is the limit of the sequence defined in Eq.~\ref{eq:seq_convergence}:  
\begin{equation} \label{eq:lim_seq}
    \begin{split}
       \mathrm{COHERENCE}(X) \approx F(\lim _{n \to \infty }v_{n}).
    \end{split}
\end{equation}

Since we cannot calculate this limit analytically, the function is defined as:

\begin{equation} \label{eq:def_coherence}
    \begin{split}
       \mathrm{COHERENCE}(X) = F(v_m) \text{ such that }  |F(v_m)-F(v_{m-1})| < \varepsilon \text{ OR } m=\mu_{\mathrm{CHR}},
    \end{split}
\end{equation}

where $\mu_{\mathrm{CHR}}$ is the maximum number of COHERENCE() iterations allowed per call. Fig.~\ref{fig:coherence} illustrates an example of one COHERENCE() call with 3 iterations. The input and first element of the sequence is $v_0=v_{\mathrm{ref}}$, and the final output is $F(v_3)$, representing the \ac{at} calculated with $v_3$, corresponding to the \ac{cv} obtained in the third iteration. This ensures that $|F(v_3)-F(v_{2})| < \varepsilon$. The function typically converges quickly, often within 2 iterations, but an extreme case close to propagation failure was chosen in the example to demonstrate the convergence process more clearly. Moreover, the example was chosen to ensure that the function $F$ has its simplest form, $t^* = a + \frac{b}{v^*}$, where $a$ is the minimum \ac{at} among the neighbors of $X$, and $b$ is the minimum possible value for $(\overrightarrow{X_{\lambda}X}^\top \mathbf{D}^{-1} \overrightarrow{X_{\lambda}X})^{\frac{1}{2}}$ in Eq.~\ref{eq:lat_triangle}
 among all triangles in $\mathcal{T}$ containing $X$ and all $\lambda \in [0,1]$ (Fig.~\ref{fig:coherence}a). If $\mathbf{D}$ is the identity matrix (i.e, in the isotropic case), then $b$ is the minimum Euclidean distance between $X$ and the perimeter of the polygon formed from the union of all triangles containing $X$. In this example, the function $F$ is monotonically decreasing, continuous, and differentiable within $(0, +\infty)$. For this to occur, the neighboring node with the minimum \ac{at} must belong to the triangle where the minimum $(\overrightarrow{X_{\lambda}X}^\top \mathbf{D}^{-1} \overrightarrow{X_{\lambda}X})^{\frac{1}{2}}$ is found. In the general case where this condition is not necessarily met, the function $F$ remains monotonous and continuous but may not be differentiable everywhere. Additionally, $F(v^*)$ tends to $+\infty$ as $v^*$ approaches $0$, and $F(v^*)$ tends to $a$ as $v^*$ approaches $+\infty$ for any situation.

The proof of convergence of the sequence described in Eq.~\ref{eq:seq_convergence} is based on the idea that the sequence can be partitioned into 2 converging subsequences: one increasing ($v_{2n}$) and the other decreasing ($v_{2n+1}$). Furthermore, these subsequences are bounded between $v_0$ and $v_1$, and the difference between $v_n$ and $v_{n+1}$ diminishes as $n$ approaches infinity.  In some cases, the sequence might need to take values of the \ac{di} below the parameter $\theta$ before converging. Therefore, when computing COHERENCE(), \ac{di} values below $\theta$ are allowed as long as the final \ac{di} is larger than $\theta$, otherwise $\mathrm{COHERENCE}(X) = \infty$. 

\subsubsection{Approximation of the Diffusion Current} \label{subsubsec:Idiff}
To allow for $V_\mathrm{m}$ calculation in coarse meshes, a current $I_{\mathrm{diff}}$ was introduced following the idea of the \ac{re}  model~\cite{neic2017efficient}. $I_{\mathrm{diff}}$ is computed in step~$\mathrm{B}_n$ and approximates $\nabla \cdot (\boldsymbol{\sigma}_{\mathrm{m}}\nabla V_{\mathrm{m}})$ in Eq.~\ref{eq:monodomain}. A similar approach is used in the eikonal model by Gassa et al.~\cite{gassa2021spiral}. In the \ac{dream}, $I_{\mathrm{diff}}$ is defined as a triple Gaussian function:

\begin{equation} \label{eq:diff_fit}
\nabla \cdot (\boldsymbol{\sigma}_{\mathrm{m}} \nabla V_{\mathrm{m}}) \Big|_{X} \approx I_{\mathrm{diff}}(X) = \sum_{i=1}^{3} \alpha_i \exp\left(-\frac{(t - T(X) - \beta_i)^2}{\gamma_i^2}\right),
\end{equation}

where the parameters $\alpha_i$, $\beta_i$ and, $\gamma_i$ are optimized to reduce the difference between $I_{\mathrm{diff}}$ and the diffusion current specific to the used ionic model. A high-resolution mesh (i.e., average edge length 200\,µm) was used to run a monodomain simulation in the optimization process. A planar wavefront was simulated by stimulating at the mesh border. The diffusion current was calculated at a node located far from the boundaries and the stimulus location. $I_{\mathrm{diff}}$ serves as a trigger for an \ac{ap}, regardless of whether the activation is initiated directly by a stimulus or through diffusion from neighboring nodes.

\subsubsection{Approximated Reaction Diffusion Model}\label{subsubsec:approxRD}
After the $n$-th \ac{cycfim} cycle is finished and $I_{\mathrm{diff}}$ is computed, the approximated \ac{rd} model will run (i.e, step~$\mathrm{C}_n$). The time interval for this step is: $[t_{\mathrm{min,init}}-\tau_{\mathrm{s}},\mathrm{min}(t_{\mathrm{min}}-\tau_{\mathrm{s}},t_{\mathrm{end}})]$ ensuring that the nodes that trigger an \ac{ap} during this step have converged \acp{at}. The diffusion term in the parabolic equation of the \ac{rd} system is replaced by $I_{\mathrm{diff}}$:
\begin{equation} \label{eq:dream}
\beta C_{\mathrm{m}} \frac{\partial V_{\mathrm{m}}}{\partial t} = I_{\mathrm{diff}} -\beta I_{\mathrm{ion}}(V_{\mathrm{m}},\overrightarrow{\eta}).
\end{equation}
In this case, the diffusion current is approximated by $I_{\mathrm{diff}}$ and only the reaction (ODE) part of the \ac{rd} system remains to be solved. Solving this ODE system independently for each node in $\Omega_{\mathcal{T}}$ will yield the values for $V_{\mathrm{m}}$ within the interval $[t_{\mathrm{min,init}}-\tau_{\mathrm{s}},\mathrm{min}(t_{\mathrm{min}}-\tau_{\mathrm{s}},t_{\mathrm{end}})]$.

\subsubsection{Identification of Repolarization Times}\label{subsubsec:findRT}
Before the $(n+1)$-th \ac{cycfim} cycle starts, it is important to identify which nodes and when are available for reactivation. Ideally, $V_{\mathrm{m}}$ of every activated node would reach its \ac{rt} during step~$\mathrm{C}_n$. However, recently activated nodes will not have recovered at the end of this \ac{rd} iteration (i.e., at $t_{\mathrm{min}}-\tau_{\mathrm{s}}$). The aim in step~$\mathrm{D}_n$ is to compute \acp{rt} of these nodes. To obtain these times, the ODEs of the single cell models are integrated in time until the threshold $-40$\,mV is crossed:

\begin{equation} \label{eq:unicellular}
C_{\mathrm{m}} \frac{\mathrm d  V_{\mathrm{m}}}{\mathrm d t} = - I_{\mathrm{ion}}(V_{\mathrm{m}},\overrightarrow{\eta}).
\end{equation}
This process is implemented for each node that was depolarized but not yet repolarized in the previous step.
The initial conditions are defined by the state variables at $t=t_{\mathrm{min}}-\tau_{\mathrm{s}}$ for each node. After the \ac{rt} is calculated, $V_{\mathrm{m}}$ and $\overrightarrow{\eta}$ are reset to the values they had at $t=t_{\mathrm{min}}-\tau_{\mathrm{s}}$.  

\subsubsection{Considerations for Empty List Scenarios}\label{subsubsec:emptylist}
Additional considerations are required if $L$ is empty but not all the boundary conditions have been computed. Let $T_{s}$ be the time when the next stimulus will be applied, then $V_{\mathrm{m}}$ is computed until $t=T_{s}-\tau_{\mathrm{s}}$. After identifying the \acp{rt}, the stimulus at $T_{s}$ must be computed before iterating \ac{cycfim} again as detailed in  Section~\ref{subsubsec:cycfim}. If $L$ is not empty after computing the stimulus, \ac{cycfim} iterates until $t_{\mathrm{min}}>T_{s}+\tau_{\mathrm{inc}}$. Given that the list was empty at the end of the previous cycle, it is no longer feasible to establish $t_{\mathrm{min,init}}$. Hence, $T_{s}$ is employed as the reference $t_{\mathrm{min,init}}$ to monitor the increment of $t_{\mathrm{min}}$. If all nodes in $\Gamma_s$ (i.e., the region in which the $s$-th stimulus is applied) are in the refractory period at time $T_s$, the list will remain empty. In that case, $V_{\mathrm{m}}$ is computed until $t=T_{s+1}-\tau_{\mathrm{s}}$, where $T_{s+1}$ is the time of the $(s+1)$-th stimulus. If necessary, subsequent boundary conditions (i.e, stimuli applied at times $T_{s+2},\ldots,T_{n_s}$) are computed until the list $L$ gain nodes or $t=t_{\mathrm{end}}$. 

\section{Benchmarking}\label{subsec:benchmark}
To assess the \ac{dream}, it was compared with the monodomain model in 3 numerical experiments.
First, multiple simulations were run on a 2D rectangular tissue mesh, each with 2 consecutive planar wavefronts per \ac{pcl} (i.e., interval between wavefronts). The \ac{pcl} was varied across simulations to analyze \ac{cv} restitution curves. 
In the second numercial experiment, functional reentry was examined in absence of structural abnormalities, i.e. their occurrence attributed solely to slow \acp{cv} and short \acp{erp}. These reentries were induced using an S1-S2 protocol in a 2D square mesh. In the third numerical experiment, the \ac{peerp} protocol was used to investigate reentry in a realistic geometry of a left atrium~\cite{azzolin2021reproducible}. These 3 numerical experiments were conducted using meshes composed of triangular elements with 4 different resolutions. The average edge lengths were 200, 400, 800, and 1600\,µm. To facilitate the comparison between different meshes, each node in the lower resolution meshes was mapped to a corresponding node at the same position in the higher resolution meshes. To execute monodomain simulations, tissue conductivities were adjusted to match the \ac{cv} values desired for each scenario~\cite{openCARP-paper, costa2013automatic}. Anisotropy ratios were set to 4 and 2 for tissue conductivity and \ac{cv} respectively. The Courtemanche et al. model with standard parameters was used to represent the healthy ionic behavior of atrial cardiomyocytes in the three numerical experiments~\cite{courtemanche1998ionic}. On the other hand, \ac{af} was modeled by modifying ion channel conductances of the Courtemanche et al. model~\cite{loewe2016modeling}.
To additionally assess the compatibility of the \ac{dream} with different ionic models, the first numerical experiment (i.e, 2D rectangular tissue mesh) was also computed utilizing the simplified ionic models of Bueno-Orovio et al.~\cite{bueno2008minimal} and Mitchell \& Schaeffer~\cite{mitchell2003two}. These ionic models were originally developed for ventricles. Therefore, their parameters were adapted for atrial tissue to reflect cellular behavior of healthy and \ac{af} remodeled cardiomyocytes. The Bueno-Orovio et al. model was modified as proposed by Lenk et al.~\cite{lenk2015initiation} providing parameter sets for both conditions. The Mitchell \& Schaeffer ionic model was adjusted according to Gassa et al.~\cite{gassa2021spiral} for healthy atrial cardiomyocytes and case 3 in Corrado et al.~\cite{corrado2016personalized} for \ac{af} remodeled cardiomyocytes.
 
\ac{cv} restitution curve parameters $\rho$, $\kappa$, and $\theta$ (Eq.~\ref{eq:function_cv}) were tuned to align the \acs{cv} restitution curve of the \ac{dream} with either healthy or \ac{af} in monodomain simulations. This adjustment was based on the \ac{cv} restitution curves obtained from monodomain simulations in a mesh with an average edge length of 200\,µm, (i.e, high resolution mesh) with 2 planar wavefronts resulting in the parameters in Table~\ref{tab:cv_restitution}.

The Crank-Nicolson method was used to solve the parabolic equation of the monodomain model. Moreover, the methods described in Section~\ref{subsec:dream} were used to solve the eikonal equation and approximate the diffusion current in the \ac{dream}. The ODEs of the ionic model were solved using the forward Euler method for $V_{\mathrm{m}}$ and the Rush-Larsen method for the gating variables in both the monodomain model and the \ac{dream}. 

Slab geometries were created using Gmsh~\cite{geuzaine2009gmsh} and processed with ParaView~\cite{ParaView}, Meshmixer~\cite{10.1145/1837026.1837034} and MeshLab~\cite{LocalChapterEvents:ItalChap:ItalianChapConf2008:129-136}. openCARP~\cite{openCARP-paper} was used to run simulations with both the \ac{dream} and monodomain model. Plots were generated using the Python library matplotlib~\cite{Hunter:2007}.

\subsection{Multiple Stimulations and CV Restitution}\label{subsec:prepacing}
To assess the \ac{dream}'s ability to faithfully replicate \ac{cv} restitution, we initially examined its capacity to simulate paced beats across a range of frequencies specific for each condition. These experiments were conducted in a rectangular mesh of 50\,mm $\times$ 10\,mm $\times$ 0\,mm,  with the centroid of the mesh located at (0\,µm, 0\,µm, 0\,µm) for each of the 4 resolutions. Preferential cardiomyocyte orientation was set as constant everywhere in the geometry and aligned with the $x$-axis (i.e., $\overrightarrow{l}=(1,0,0)$). Each experiment involved pre-pacing at a single-cell level for 50 activation cycles at a cycle length of 250\,ms. For monodomain simulations, conductivities were tuned to match the target \ac{cv} of 1000\,mm/s.

\begin{table}
    \centering
    \caption{\Acl{cv} restitution parameters for function H() in healthy and \acl{af} (\ac{af}) cases for ionic models by: Courtemanche et al., Bueno-Orovio et al. and Mitchell \& Schaeffer, used in the \ac{dream} simulations.}
    \begin{tabular}{ccccccc}
    \toprule
     & \multicolumn{2}{c}{Courtemanche et al.} & \multicolumn{2}{c}{Bueno-Orovio et al.} & \multicolumn{2}{c}{Mitchell \& Schaeffer} \\ 
     \cmidrule(l){2-3}\cmidrule(l){4-5}\cmidrule(l){6-7}
       Parameter & Healthy & \ac{af} & Healthy & \ac{af} & Healthy & \ac{af}\\
    \midrule
        $\rho~\mathrm{(-)}$          & 811     & 100000 & 55.13     & 100000 & 4.41     & 2.90\\
    
        $\kappa~\mathrm{(ms)}$          & 53      & 116.41 & 138.14     & 118.16 & 161.47     & 151.75\\
    
        $\theta~\mathrm{(ms)}$          & 137     & 66.28  & 139.73     & 70.64 & 129.30     & 103.12\\
    
        $\psi~\mathrm{(ms)}$          & 159     & 159  & 159     & 159 & 159     & 159\\
    \bottomrule
    \end{tabular}
    \label{tab:cv_restitution}
\end{table}

In this numerical experiment, 2 planar wavefronts were generated for each \ac{pcl} by stimulating the left side of the mesh with different frequencies. For monodomain simulations, the stimulus was applied as a transmembrane current density with an amplitude of 300\,µA/cm$^{\mathrm{2}}$ for a duration of 2\,ms. For the healthy atrial electrophysiology condition with the ionic model of Courtemanche et al., the \ac{pcl} was decreased in 50\,ms intervals from 950\,ms to 400\,ms, then in 10\,ms intervals from 390\,ms to 320\,ms, and finally in 1\,ms intervals from 320\,ms to 318\,ms. The different intervals were chosen to cover critical changes in electrophysiological behavior, as \acs{cv} and \ac{erp} drastically decrease towards the last propagating \ac{pcl}. For the ionic models of Bueon-Orovio et al. and Mitchell \& Schaeffer, also shorter \ac{pcl}s propagated on the mesh. Therefore, the \ac{pcl} was further decreased in 1\,ms intervals to 250\,ms. In the \ac{af} condition, the \ac{pcl} was decreased for all ionic models from 950\,ms to 200\,ms in 50\,ms intervals, and from 199\,ms to 118\,ms in 1\,ms intervals. The considered \ac{pcl} values varied across conditions due to differences in the restitution curves, requiring a range of \ac{pcl} values tailored to each condition.

After the experiments, the \ac{cv} restitution curves were determined. To calculate the \ac{cv}, the \acp{at} of the mesh points (--15000\,µm, 0\,µm, 0\,µm) and (15000\,µm, 0\,µm, 0\,µm) were used. \ac{cv} was normalized over $v_{\mathrm{ref}}$. The \ac{erp} was defined as the time difference between \acp{at} and \acp{rt} from the second activation cycle plus $\theta$ from Eq.~\ref{eq:function_cv} (i.e., shortest propagating \ac{di}).

\subsection{Reentry in 2D Slabs}\label{subsubsec:reentry2d}
To evaluate the impact of mesh resolution on reentry properties in both models, a 2D slab geometry was subjected to S1-S2 protocols. The geometry was represented by squared unstructured meshes of size 51.2\,mm $\times$ 51.2\,mm $\times$ 0, with the centroid of the mesh located at (25600\,µm, 25600\,µm, 0\,µm) for each of the 4 resolutions. Preferential cardiomyocyte orientation was set as constant everywhere in the geometry and aligned with the $x$-axis (i.e., $\overrightarrow{l}=(1,0,0)$). Experiments were conducted for homogeneous reference \ac{cv} values of 200, 600 and 1000\,mm/s.
The  Courtemanche et al. ionic model was pre-paced at 250\,ms \ac{pcl} for 50 activation cycles.
The tissue was pre-paced 5 times with planar stimuli (S1) from the left border of the slab. Subsequently, a single cross-field stimulus (S2) was applied in the bottom left quadrant at various times to induce reentry. After each simulation, it was assessed whether S2 was applied too early and got completely blocked, applied too late and propagated to all directions without any block, or if unidirectional block occurred as a prerequisite to induce reentry. Reentry duration was defined as the period from S2 application to the last \ac{at} in the simulation. $\Delta S$ was defined as the difference between the times when S2 and the last S1 were applied. The vulnerable window duration was determined as the difference between the earliest and latest $\Delta S$ values that induced reentry. The sample frequency for $\Delta S$ values was 1\,ms. The mean and standard deviation of the local reentry cycle length were calculated at 4 nodes, each located at the same coordinates in all 4 mesh resolutions: $P_{\mathrm{left}}$ = (12800\,µm, 25600\,µm, 0\,µm), $P_{\mathrm{down}}$ = (25600\,µm, 12800\,µm, 0\,µm), $P_{\mathrm{right}}$ = (38400\,µm, 25600\,µm, 0\,µm), and $P_{\mathrm{up}}$ = (25600\,µm, 38400\,µm, 0\,µm). To calculate the local reentry cycle length, reentries were induced using \ac{dream} and monodomain simulations with a reference conduction velocity (\ac{cv}) of 200\,mm/s and a stimulus interval ($\Delta S$) of 225\,ms in the 4 resolutions. The local reentry cycle length was defined as the difference between the \acp{at} corresponding to 2 consecutive activation cycles of the same node after reentry was induced. 

\subsection{Reentry in the Left Atrium}\label{subsubsec:reentry3d_m}

The \ac{peerp} method~\cite{azzolin2021reproducible} was used to induce reentries in a realistic human left atrial geometry from a publicly available dataset~\cite{azzolin_2021_5589289}. This realistic geometry is derived from an instance of a statistical shape model~\cite{nagel2021bi}. Preferential cardiomyocyte orientation was assigned using a rule-based method~\cite{azzolin2023augmenta}. The endocardial surface was extracted and remeshed at 4 different resolutions to achieve the desired average edge lengths. To benchmark the \ac{dream} on coarse meshes, the adaptations detailed below were introduced; the remaining parameters were sourced from Azzolin et al.~\cite{azzolin2021reproducible}. Modifications were applied equally to both \ac{dream} and monodomain experiments. From the coarsest mesh (1600\,µm average edge length), 21 points that were evenly spaced approximately 2\,cm apart were selected. Since the lower resolution meshes were embedded in the higher resolution ones, each of the 21 points were located at positions where each of the 4 meshes had a corresponding node. The reference \ac{cv} was reduced to 200\,mm/s, and the conductances of the ionic model were adapted to represent \ac{af} to increase the likelihood of inducing reentry without the need for an additional arrhythmogenic substrate. Pre-pacing on single cell and tissue level was performed for 50 and 5 activation cycles, respectively, at a \ac{pcl} of 250\,ms. From the \acp{ap} triggered by the last tissue pre-pacing, the \ac{rt} was calculated and the \ac{erp} was defined as \ac{rt}$+\theta$. Afterwards, the tissue was stimulated at the end of the \ac{erp} to try to induce reentry at each of the 21 chosen locations. Stimulation comprised all nodes within a 5\,mm-radius sphere centered at this location. Stimulations were stopped if a reentry was induced or the maximum number of 4 stimuli per location was reached.

\subsection{Computing Time}\label{subsec:comptimes_m}

The simulations were executed on a 2017 iMac, equipped with a 3.5\,GHz Quad-Core Intel Core i5 processor and 64\,GB 2400\,MHz DDR4 memory. Computing times were benchmarked in 2 experiments across 4 different resolutions using Courtemanche et al. as ionic model. In the first experiment, one single planar wavefront was stimulated in each of the 2D slabs described in Section~\ref{subsubsec:reentry2d}. The computing time was measured from the onset of the stimulus until 500\,ms of simulated time thereafter. The selected duration encompassed the repolarization phases of all nodes. In the second experiment, a reentry was induced using the S1-S2 protocol presented in Section~\ref{subsubsec:reentry2d}. The computing time was measured starting 200\,ms of simulated time after delivering the S2 stimulus and ending 300\,ms of simulated time after delivering the S2 stimulus, i.e., in the interval [S2+200\,ms, S2+300\,ms]. 
\FloatBarrier

\section{Results} \label{sec:results}

\subsection{Multiple Stimulation and CV Restitution}\label{subsec:prepacing_r}
 \begin{figure}
 \centering
\includegraphics[width=120mm]{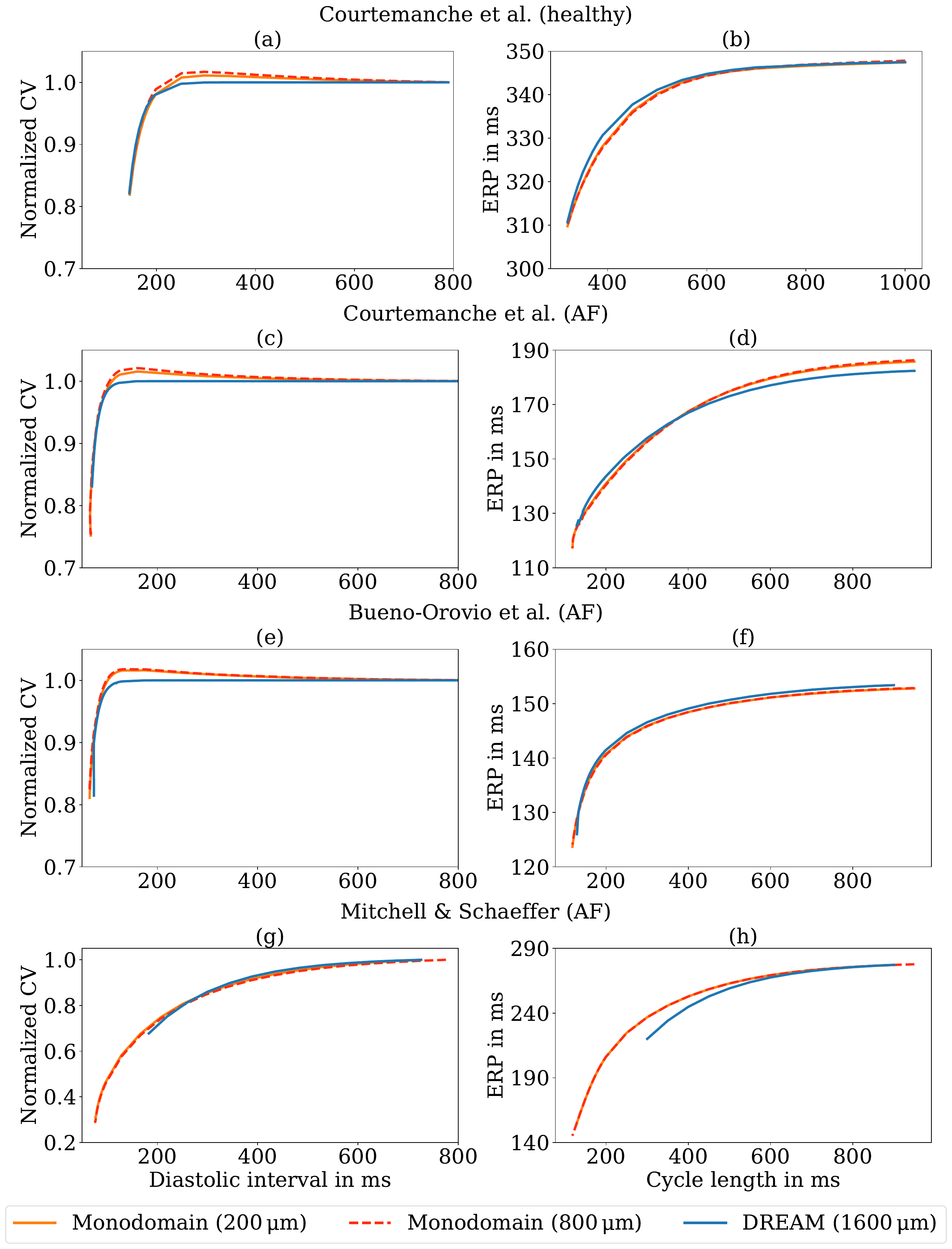}
\caption{Restitution curves for conduction velocity (\ac{cv}) and effective refractory period (\ac{erp}) obtained from experiments using meshes with average element edge lengths of 200\,µm and 800\,µm for the monodomain model, and 1600\,µm for the \ac{dream}, with the healthy and \ac{af} remodeled Courtemanche et al. ionic model, as well as the \ac{af} remodeled ionic models of Bueno-Orovio et al. and Mitchell \& Schaeffer. Results for the monodomain model with an edge length of 400\,µm are not shown due to negligible differences compared to the 200\,µm results, and the monodomain model with 1600\,µm edge length is excluded due to propagation failure. Finer resolutions for the \ac{dream} (i.e., 200\,µm, 400\,µm, and 800\,µm) are not displayed as they are mostly indistinguishable from the 1600\,µm results. \textbf{a, c, e, g)} \ac{cv} restitution curves normalized to the reference \ac{cv}, $v_{\mathrm{ref}} = 1000\,\mathrm{mm/s}$ and \textbf{b, d, f, h)} \ac{erp} restitution curves for the Courtemanche et al. ionic model representing healthy and \ac{af} remodeled behavior, as well as the \ac{af} remodeled ionic models of Bueno-Orovio et al. and Mitchell \& Schaeffer, respecitively.}
\label{fig:restCV}
\vspace{-10pt}
\end{figure}

Fig.~\ref{fig:restCV} shows the comparison of \ac{cv} and \ac{erp} restitution curves between the \ac{dream} and monodomain models at resolutions of 200, 400, 800, and 1600\,µm. The analysis reveals distinct characteristics for both the healthy and AF cases. Specifically, the ionic model of Courtemanche et al. was used for the healthy and AF remodeled case, whereas the AF remodeled case employed the ionic models of Bueno-Orovio et al. and Mitchell \& Schaeffer. Table~\ref{tab:rmse_cverp_restitution} shows the \ac{rmse} of \ac{cv} and \ac{erp} restitution curves for all considered ionic models with both healthy and \ac{af} conditions. Monodomain simulations on meshes with an average edge length 200\,µm were used as reference solution to calculate \ac{rmse}. Addressing simulations with the Courtemanche et al. ionic model, \ac{dream} and monodomain model exhibited similar steepness across all resolutions for the healthy case in both restitution curves (Fig.~\ref{fig:restCV}a and Fig.~\ref{fig:restCV}b), with \acp{rmse} below 8.85\,mm/s and 1.52\,ms respectively.  In the \ac{af} case (Fig.~\ref{fig:restCV}c and Fig.~\ref{fig:restCV}d), both propagation models maintained similar steepness and shortest propagating \ac{di} across resolutions. However, the \ac{dream} notably did not represent the temporary raise of \ac{cv} during intermediate \acp{di}, while the monodomain model showed varying levels of biphasic restitution across resolutions, with lower resolutions having a higher maximum \ac{cv} at intermediate \acp{di}. The monodomain model demonstrated smaller last propagating cycle lengths, indicating the shortest possible \acp{di} before propagating failure occurred. A more discernible difference arose in the \ac{cv} resulting from each propagation model at this minimal \ac{di}. The \ac{dream} demonstrated a higher minimum \ac{cv} compared to the monodomain model (Fig.~\ref{fig:restCV}c). 
The ranges of both restitution curves obtained with \ac{dream} and monodomain model for the Bueno-Orovio et al. ionic model resemble those for the Courtemanche et al. ionic model. The \ac{cv} and \ac{erp} restitution curves (shown for the \ac{af} case in Fig.~\ref{fig:restCV}e and Fig.~\ref{fig:restCV}f) showed similar steepness. Similar to simulations with the \ac{dream} and Courtemanche et al., the biphasic \ac{cv} restitution obtained with the monodomain model and Bueno-Orovio et al. was not reproduced with the \ac{dream} and Bueno-Orovio et al. 
The restitution curves for the ionic model of Mitchell \& Schaeffer (shown for the \ac{af} remodeled case in Fig.~\ref{fig:restCV}g) and Fig.~\ref{fig:restCV}h)
showed higher last propagating cycle lengths for the \ac{dream} compared to the monodomain model. Pacing cycle length and \ac{di} for the last possible propagation were about 173\,ms and 107\,ms, respectively, longer for the \ac{dream} compared to the monodomain models. Despite the absence of biphasic behavior in the monodomain \ac{cv} restitution curve, the \ac{rmse} values were consistently higher across resolutions for the \ac{dream} compared to Courtemanche et al. ionic model. 
Monodomain simulations resulted in smaller errors for both restitution curves under both conditions for resolutions without propagation failure. Moreover, \ac{dream} simulations showed a tendency towards higher errors for ionic models with reduced levels of detail. Simulations across different resolutions for the \ac{dream} showed similar \acp{rmse} for each condition and ionic model. For instance differences in \acp{rmse} among resolutions in \ac{dream} simulations were smaller than 5.46\,mm/s and 0.5\,ms for the \ac{cv} and \ac{erp} restitution curves respectively per condition and ionic model. 
 
\begin{table}[h!]
\centering
\caption{Root mean square error of conduction velocity (CV) restitution and effective refractory period (ERP) restitution for healthy and \acl{af} (\ac{af}) conditions, in \ac{dream} and monodomain model with Courtemanche et al., Bueno-Orovio et al., and Mitchell \& Schaeffer embedded ionic models at various resolutions indicated by average edge length, compared to the restitution obtained with the monodomain model at 200\,µm.}

\begin{tabular}{@{\extracolsep\fill}llccccc}
\toprule
&\multicolumn{2}{@{}c@{}}{}& \multicolumn{2}{@{}c@{}}{\ac{cv} (mm/s)} & \multicolumn{2}{@{}c@{}}{\ac{erp} (ms)}
\\\cmidrule(r){4-5}\cmidrule(l){6-7}%
Ionic model & Propagation Model & Resolution (µm) & Healthy& \ac{af} & Healthy  & \ac{af}  \\
\midrule
\multirow{8}{*}{Courtemache et al.} & \multirow{4}{*}{\ac{dream}} & 200 & 8.85 & 7.19 & 1.52 & 2.39 \\
& & 400 & 8.74 & 7.14 & 1.52 & 2.39 \\
& & 800 & 8.47 & 7.14 & 1.50 & 2.39 \\
& & 1600 & 8.01 & 7.26 & 1.49 & 2.38 \\
\cmidrule(r){2-7}
& \multirow{4}{*}{Monodomain\footnotemark[1] } & 200 & 0.00 & 0.00 & 0.00 & 0.00 \\
& & 400 & 0.63 & 2.60 & 0.36 & 0.09 \\
& & 800 & 5.88 & 5.17 & 0.25 & 0.46 \\
& & 1600 & (-) & (-) & (-) & (-) \\
\midrule
\multirow{8}{*}{Bueno-Orovio et al.} & \multirow{4}{*}{\ac{dream}} & 200 & 7.21 & 15.95 & 1.03 & 1.11 \\
& & 400 & 6.62 & 13.60 & 1.09 & 1.12 \\
& & 800 & 6.48 & 11.49 & 1.02 & 1.12 \\
& & 1600 & 7.07 & 12.08 & 1.05 & 1.17 \\
\cmidrule(r){2-7}
& \multirow{4}{*}{Monodomain\footnotemark[1]} & 200 & 0.00 & 0.00 & 0.00 & 0.00 \\
& & 400 & 0.98 & 0.30 & 0.49 & 0.08 \\
& & 800 &  14.36 & 4.13 & 1.53 & 0.18 \\
& & 1600 & (-) & (-) & (-) & (-) \\
\midrule
\multirow{8}{*}{Mitchell \& Schaeffer} & \multirow{4}{*}{\ac{dream}} & 200 & 17.36 & 10.82 & 7.09 & 6.38 \\
& & 400 & 18.21 & 8.01 & 7.21 & 6.44 \\
& & 800 & 17.65 & 7.03 & 7.22 & 6.46 \\
& & 1600 & 19.28 & 7.76 & 7.24 & 6.46 \\
\cmidrule(r){2-7}
& \multirow{4}{*}{Monodomain\footnotemark[1]} & 200 & 0.00 & 0.00 & 0.00 & 0.00 \\
& & 400 & 2.80 & 0.30 & 0.54 & 0.03 \\
& & 800 & 2.20 & 5.33 & 0.52 & 0.04 \\
& & 1600 & (-) & (-) & (-) & (-) \\
\bottomrule
\end{tabular}
\footnotetext[1]{Propagation failure at 1600\,µm average edge length.}
\label{tab:rmse_cverp_restitution}
\end{table}

\FloatBarrier
\subsection{Reentry in 2D Slabs}\label{subsec:reentry2d_r}

\begin{figure}
\centering
\includegraphics[width=120mm]{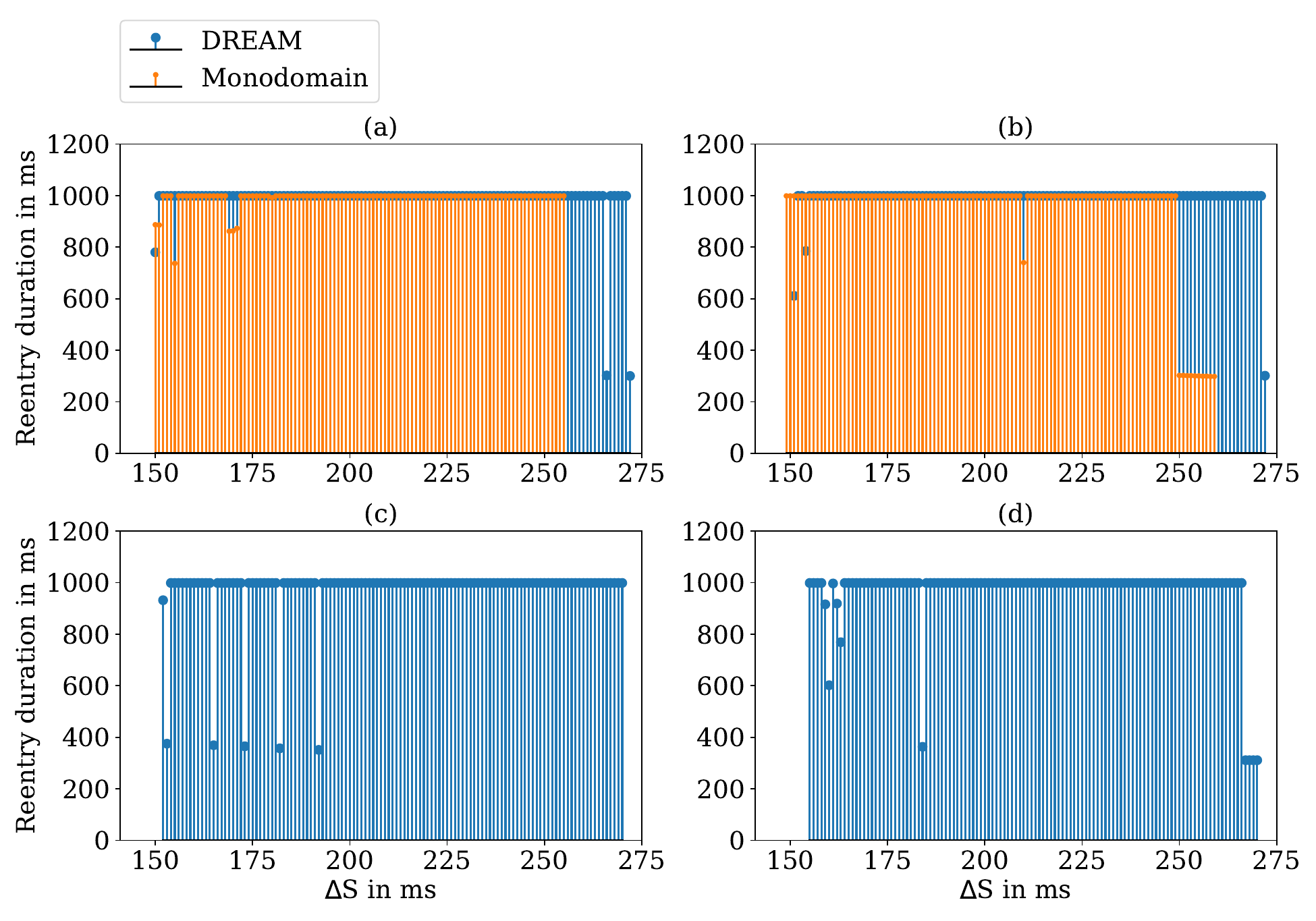}
\caption{Reentry duration at a conduction velocity of 200\,mm/s for different mesh resolutions. Average edge length: \textbf{a)} 200\,µm, \textbf{b)} 400\,µm, \textbf{c)} 800\,µm, \textbf{d)} 1600\,µm. Monodomain simulations for coarser resolutions are not shown due to propagation failure in \textbf{c)} and \textbf{d)}). The horizontal axis shows $\Delta S$ (time difference between S2 and S1). The vertical axis shows reentry duration.
}
\label{fig:reentry200}
\vspace{-10pt}
\end{figure}

 Fig.~\ref{fig:reentry200} shows the reentry duration for $\Delta S$ values within the vulnerable window for each of the resolutions at a \ac{cv} of 200\,mm/s. The \ac{dream} produced similar results across all resolutions. In contrast, the monodomain results failed to propagate during pre-pacing for coarser resolutions (800 and 1600\,µm average edge length). For higher resolutions (200 and 400\,µm average edge length), both models showed reentries that lasted 1000\,ms (i.e., until the simulation reached $t_{\mathrm{end}}$) for most of the tested $\Delta S$ values. However, monodomain simulations produced a few reentries that stopped before 1000\,ms for less than 10 $\Delta S$ values sparsely distributed across the vulnerable window. Moreover, vulnerable windows were shorter for the monodomain model mainly because it did not induce reentries for the longest $\Delta S$ values. For higher \acp{cv}, the monodomain model produced reentries that terminated before the end of the simulation, whereas simulations with the \ac{dream} yielded a unidirectional block but did not reactivate the nodes where the S2 stimulus was applied. 
 
\begin{figure}
\centering
\includegraphics[width=120mm]{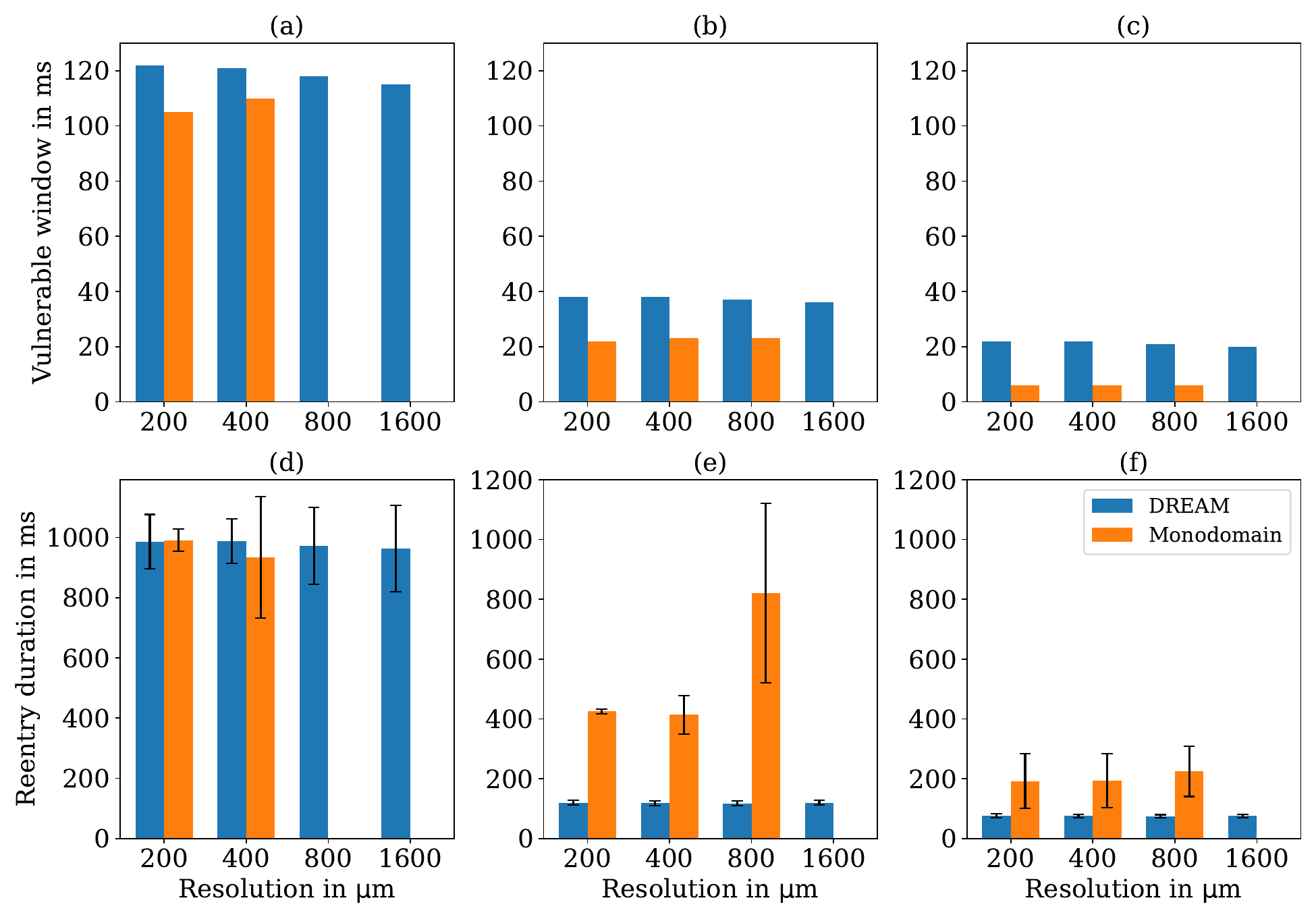}
\caption{(\textbf{a}) Vulnerable window at different mesh resolutions for a conduction velocity (CV) of 200\,mm/s, (\textbf{b}) 600\,mm/s  and (\textbf{c}) 1000\,mm/s. (\textbf{d}) Reentry duration (mean$\pm$std) for a CV of 200\,mm/s, (\textbf{e}) 600\,mm/s, (\textbf{f}) and 1000\,mm/s.}
\label{fig:reentry_vulnerabilitywindow}
\vspace{-10pt}
\end{figure}

Fig.~\ref{fig:reentry_vulnerabilitywindow}a), Fig.~\ref{fig:reentry_vulnerabilitywindow}b) and Fig.~\ref{fig:reentry_vulnerabilitywindow}c) show the vulnerable window duration for the 2 models at each of the 4 resolutions at different reference \acp{cv}. The vulnerable window duration was defined as the time between the earliest and the latest S2 time that induced a unidirectional block. Monodomain simulations on the mesh with an average edge length of 200\,µm were taken as ground truth. The \ac{dream} kept a stable error as mesh resolution decearse, performing equally even at the coarsest resolution of 1600\,µm. Vulnerable windows were longer for all \acp{cv} and all the resolutions in the \ac{dream} simulations. These differences in vulnerable window duration between \ac{dream} simulations in all resolutions and monodomain simulation at finer resolutions did not increase in simulations with higher reference \ac{cv}. Fig.~\ref{fig:reentry_vulnerabilitywindow}d), Fig.~\ref{fig:reentry_vulnerabilitywindow}e) and Fig.~\ref{fig:reentry_vulnerabilitywindow}f) show the reentry duration (mean$\pm$std) for the same reference \acp{cv}. For the slowest \ac{cv} both models produced longer reentries at most of $\Delta S$ values. In faster \ac{cv}, both models resulted in shorter durations for all resolutions. The \ac{dream} simulations resulted in reentry duration errors across all resolutions that increased with higher \acp{cv}. The discrepancy is grounded in the fact that the monodomain simulations produced self-terminated shorter reentries with a few turns within the slab at higher reference \ac{cv} in this particular experiment. The \ac{dream} simulations, on the other hand, resulted in propagation patterns with unidirectional blocks that were incapable of completing a full turn due to the lack of source-sink mismatch representation.

Table~\ref{tab:reentry_cycle_length} presents the mean and standard deviation of local reentry cycle lengths observed in \ac{dream} and monodomain simulations across various resolutions at nodes $P_{\mathrm{left}}$, $P_{\mathrm{down}}$, $P_{\mathrm{right}}$, and $P_{\mathrm{up}}$. In monodomain simulations with average edge lengths of 200\,µm, mean local reentry cycle lengths ranged from 136 to 157\,ms, with standard deviations varying from 17 to 28\,ms. \ac{dream} simulations consistently showed comparable mean and standard deviation values across different resolutions, ranging from 137 to 164\,ms and from 9 to 46\,ms, respectively. Notably, most of mean local reentry cycle lengths from \ac{dream} simulations were closer to those observed in monodomain simulations at 200\,µm than in simulations at 400\,µm.

\begin{table}[htbp]
\centering
\caption{Mean $\pm$ standard deviation of the local reentry cycle length in ms at 4 points $P_x$ across various mesh resolutions, indicated in average edge lengths in µm, for both the \ac{dream} and monodomain models.}
\begin{tabular}{lcccccc|}
\toprule
Propagation Model & Resolution & $P_{\mathrm{left}}$ & $P_{\mathrm{down}}$  & $P_{\mathrm{right}}$  & $P_{\mathrm{up}}$  \\
\midrule
\multirow{4}{*}{\ac{dream}} & 200 & 154 $\pm$ 45 & 137 $\pm$ 11 & 151 $\pm$ 28 & 161 $\pm$ 45 \\

 & 400 & 156 $\pm$ 46 & 138 $\pm$ 10 & 153 $\pm$ 28 & 163 $\pm$ 44 \\

 & 800 & 158 $\pm$ 40 & 143 $\pm$ 9  & 154 $\pm$ 22 & 164 $\pm$ 38 \\

 & 1600 & 157 $\pm$ 33 & 149 $\pm$ 10 & 154 $\pm$ 20 & 162 $\pm$ 32 \\

\midrule
\multirow{4}{*}{Monodomain\footnotemark[1]} & 200 & 136 $\pm$ 28 & 157 $\pm$ 26 & 152 $\pm$ 28 & 146 $\pm$ 17 \\
 & 400 & 162 $\pm$ 15 & 174 $\pm$ 24 & 174 $\pm$ 19 & 168 $\pm$ 24 \\
 & 800 & (-) & (-) & (-) & (-) \\
 & 1600& (-) & (-) & (-) & (-) \\
\bottomrule
\end{tabular}
\footnotetext[1]{Propagation failure at 800 and 1600\,µm.}
\label{tab:reentry_cycle_length}
\end{table}

\subsection{Reentry in the Left Atrium}\label{subsec:reentry3d_r}

\begin{figure}
\centering
\includegraphics[width=120mm]{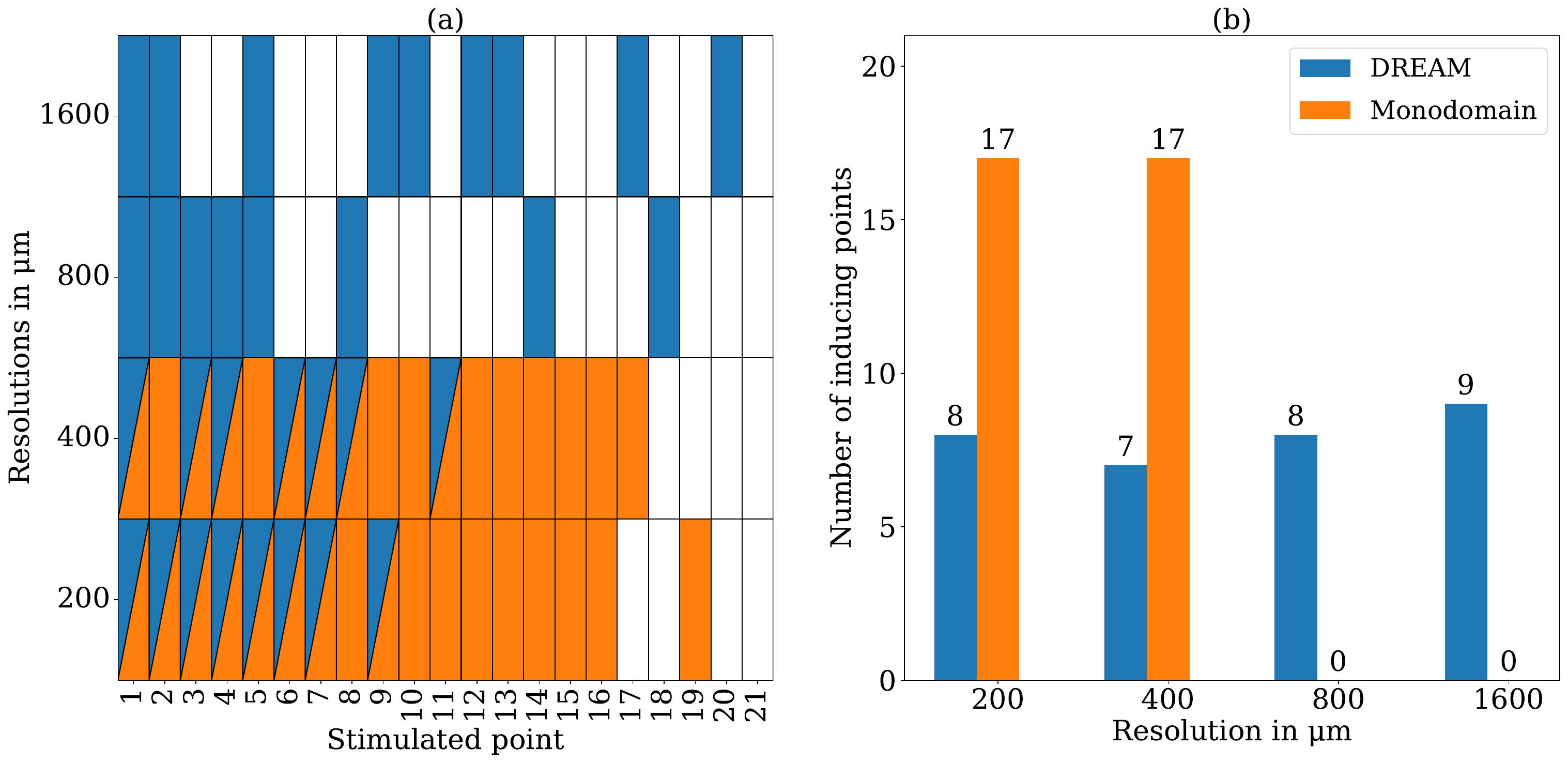}
\caption{Inducing points identified by the PEERP protocol. \textbf{a)} Distribution of points that induced reentry in experiments with the \ac{dream}, the monodomain model or both at different mesh resolutions (average edge length of 200, 400, 800, or 1600\,µm). Each row represents an experiment, and each column represents one of the 21 stimulated points. The square's color indicates points that produced reentry with the \ac{dream}, monodomain model, or both. \textbf{b)} Number of inducing points per model and resolution.}
\vspace{5pt}
\label{fig:peerp}
\end{figure}

\begin{figure}
\centering
\includegraphics[width=120mm]{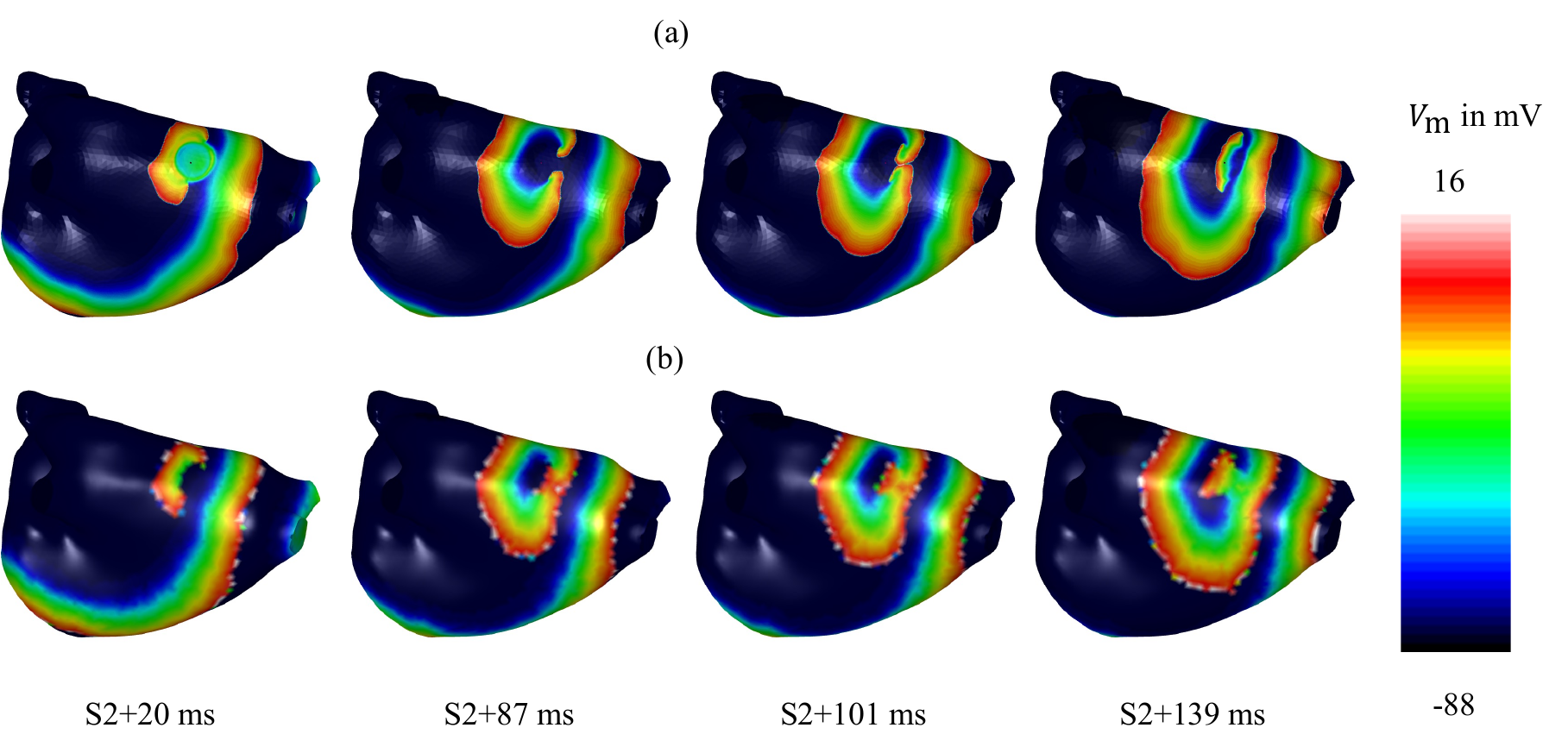}
\caption{Left atrium posterior view of transmembrane voltage maps of induced reentries at different time points after S2 , stimulated at a point close to the roof. \textbf{a)} Monodomain model with 200\,µm average edge length \textbf{b)} \ac{dream} with 1600\,µm average edge length.}
\vspace{5pt}
\label{fig:peerp_similar}
\end{figure}

\begin{figure}
\centering
\includegraphics[width=120mm]{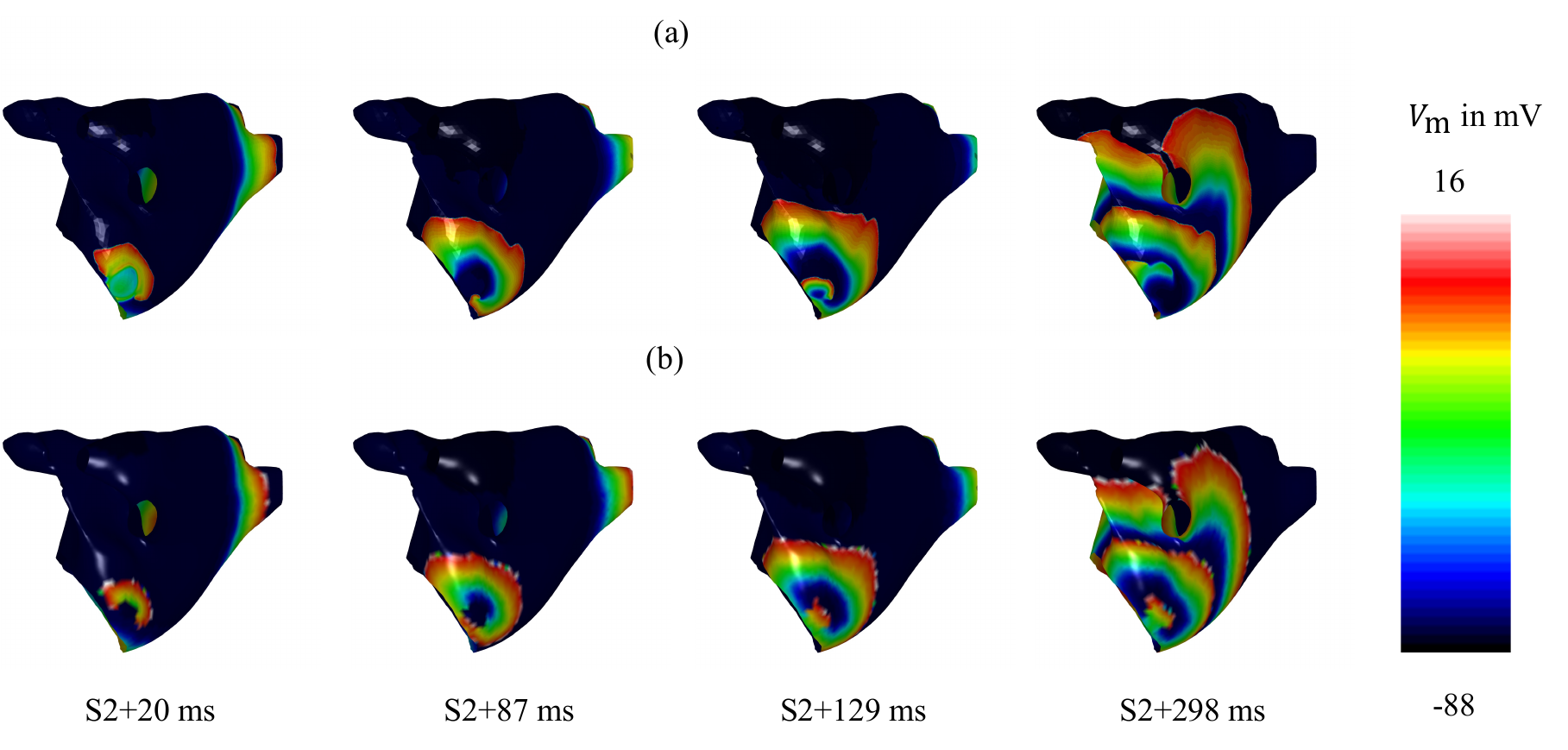}
\caption{Left atrium posterolateral view of transmembrane voltage maps of induced reentries at different time points after S2 , stimulated at a point near the mitral valve, \textbf{a)} Monodomain model with 200\,µm average edge length \textbf{b)} \ac{dream} with 1600\,µm average edge length.}
\vspace{5pt}
\label{fig:peerp_different}
\end{figure}

Fig.~\ref{fig:peerp} shows the number of inducing points of the 21 stimulation points in the left atrium per experiment. Monodomain experiments at higher resolutions (average edge lengths 200 and 400\,µm) showed the same number of inducing points. On the other hand, monodomain simulations failed to induce any reentry at lower resolutions (average edge lengths 800 and 1600\,µm) due to propagation failure. \ac{dream} experiments showed a lower number of inducing points compared to monodomain experiments. However, \ac{dream} experiments showed a similar amount of inducing points across all resolutions. For some experiments, both models showed similar mechanisms of reentry at the same inducing point. For example, Fig.~\ref{fig:peerp_similar} shows simulations in the same inducing point in which both the \ac{dream} and monodomain model produced a figure-of-eight reentry. Despite this example showing the same mechanism of reentry, the one induced by the \ac{dream} lasted longer time. Fig.~\ref{fig:peerp_different} shows an example where the \ac{dream} shows a figure-of-eight reentry while the monodomain model shows a spiral reentry. This difference arose from the lack of source-sink mismatch representation in the \ac{dream}. In both cases, the reentries were sustained for the full 1000\,ms simulated. Table~\ref{tab:sensit_specif_peerp} provides sensitivity and specificity of both models in the resolutions that did not have propagation failure running the \ac{peerp} protocol. Monodomain results at the highest resolution (i.e., average edge length of 200\,µm) were used as ground truth. \ac{dream} simulations exhibited better specificity than sensitivity, as most inducing points identified by the \ac{dream} were also found by the monodomain model, whereas not all points found by the monodomain model were captured by the \ac{dream}. On the other hand, \ac{dream} performance was consistent across all resolutions, except for a significant drop in specificity for the coarsest resolution (i.e., average edge length of 1600\,µm) to 50\,\%.
  
\begin{table}[h!]
\centering
\caption{Sensitivity and specificity for the \ac{dream} and monodomain model at different mesh resolutions, expressed in average edge length in µm, considering monodomain experiments at 200\,µm as the ground truth.}
\begin{tabular}{llcc}
\toprule
Propagation Model & Resolution & Sensitivity (\%) & Specificity (\%) \\
\midrule
\multirow{4}{*}{\ac{dream}} & 200 & 47 & 100 \\
& 400 & 41 & 100 \\
& 800 & 41 & 75 \\
& 1600 & 41 & 50 \\
\midrule
\multirow{4}{*}{Monodomain\footnotemark[1]} & 200 & 100 & 100\\
 & 400 & 94 & 75 \\
 & 800 & (-) & (-) \\
 & 1600& (-) & (-) \\
\bottomrule
\end{tabular}
\footnotetext[1]{Propagation failure at 800 and 1600\,µm.}
\label{tab:sensit_specif_peerp}
\end{table}

\subsection{Computing Times}\label{subsec:comptimes_r}

\begin{figure}
\centering
\includegraphics[width=120mm]{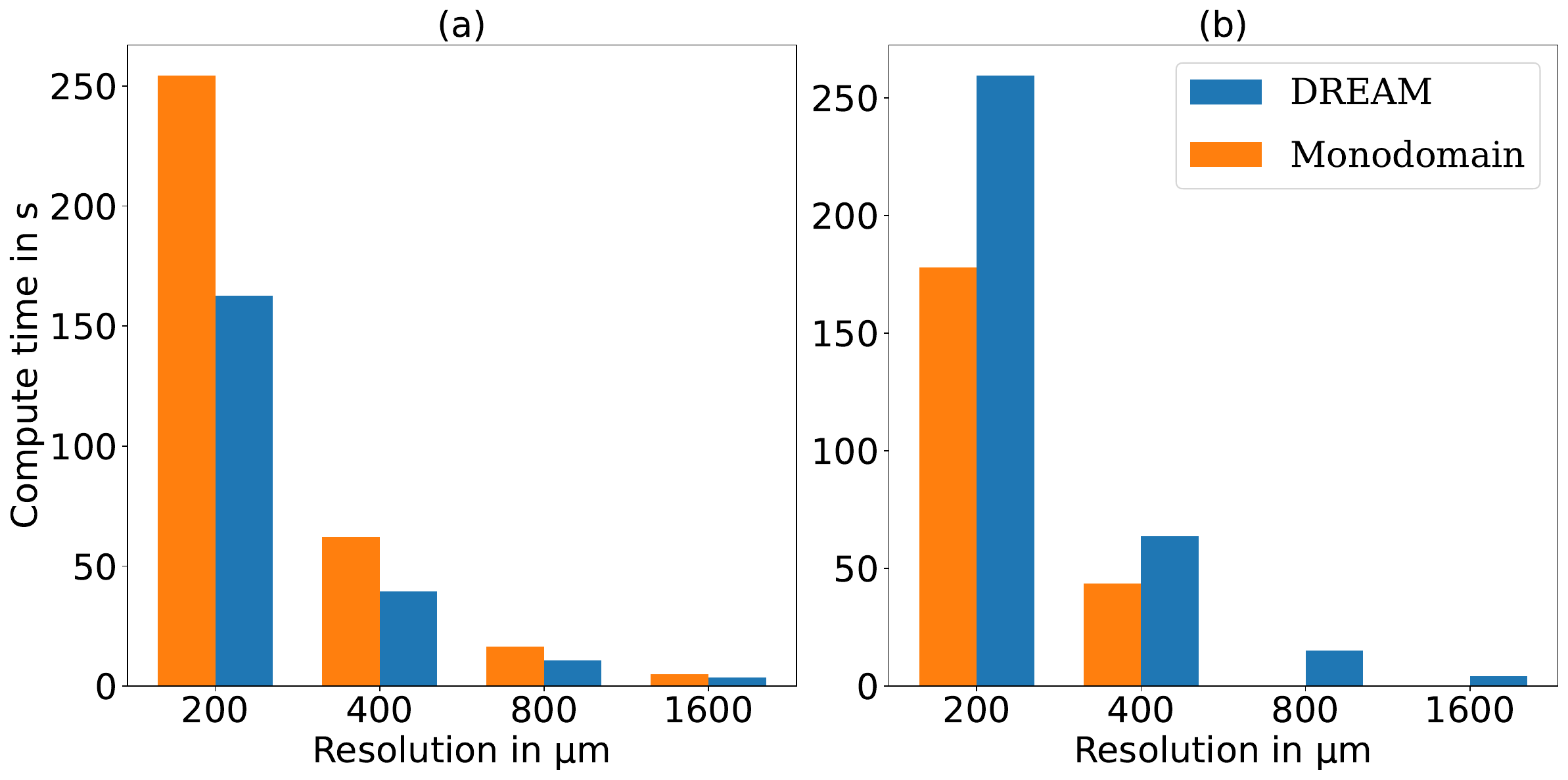}
\caption{Computing time: \textbf{a)} planar wavefront, \textbf{b)} reentry.}
\vspace{5pt}
\label{fig:comp_times}
\end{figure}

Fig.~\ref{fig:comp_times} illustrates the computing times that the \ac{dream} and monodomain model took for simulating a planar wavefront and reentry across different mesh resolutions. The proportion of computing time for the different steps of both models remained consistent across resolutions. Therefore, only the detailed computing times for each step are mentioned for simulations on a mesh with an average edge length of 200\,µm. For all resolutions, the planar wavefront experiments were faster with the \ac{dream} compared to the monodomain model. In the \ac{dream} simulations for the planar wavefront, the majority of the computing time was spent solving the ionic model equations (Fig.~\ref{fig:dream} step~C), which took 121\,s. The second most time-consuming step was iterating the \ac{cycfim} (Fig.~\ref{fig:dream} step~A), which took 26\,s. Computing $I_{\mathrm{diff}}$ (Fig.~\ref{fig:dream} step~B) required significantly less time (3.5\,s). Calculating recovery times (Fig.~\ref{fig:dream} step~D) was not necessary as there was no reactivation in this experiment. In the same scenario, the monodomain model spent nearly an equal amount of time solving the parabolic equation and computing the ionic model (130 and 123\,s respectively).

In contrast, the reentry experiments were faster with the monodomain model compared to the \ac{dream} for the same resolution across all resolutions. However, \ac{dream} simulations on coarser resolution meshes achieved approximately 40x faster computing times than monodomain simulations on finer resolution meshes. For the \ac{dream}, the majority of computing time was spent in step~A, followed by steps~C,~D and~B, taking 147, 73, 26, and 2.5\,s respectively (See Fig.~\ref{fig:dream}). Given the significant role of diffusion currents in reentries, the monodomain simulations devoted most of their time to solving the parabolic equation (101\,s), with only 74\,s spent on computing the ionic current.

\section{Discussion}\label{sec:discussion}
%

\subsection{Advantages and Novel Aspects of the \ac{dream}}\label{subsec:novel}

The \ac{dream} benefits from the consistency of the eikonal model across mesh resolutions. Across all experiments, it was possible to obtain similar results with all tested resolutions. Unlike the monodomain model that requires fine mesh resolutions, hence more computational effort, the \ac{dream} yields similar results at lower resolutions. This characteristic allows the \ac{dream} to perform faster simulations. 

Most of the eikonal models that have attempted to incorporate reentry phenomena use the \ac{fmm} as a numerical method~\cite{gassa2021spiral,pernod2011multi} and implementations are not publicly available. Moreover, the regular \ac{fmm} struggles with anisotropic propagation because this algorithm assumes the characteristic direction to be always colinear with the wavefront gradient, which is not true for anisotropic cases~\cite{sethian2003ordered}. To address this limitation, Pernod et al. developed the anisotropic \ac{fmm}, which incorporates an additional ``CHANGED'' list to enable recursive corrections~\cite{pernod2011multi}. Similarly, Cristiani proposed a similar method, introducing the ``BUFFER'' list with the same recursive correction concept~\cite{cristiani2009fast}. In his study, he conducted various tests comparing the buffered \ac{fmm} against the regular \ac{fmm}, focusing on anisotropy problems. His findings revealed that the regular \ac{fmm} calculated incorrect solutions with varying degrees of error across different anisotropic scenarios and resolutions. For instance, in an experiment with an anisotropy ratio of approximately 5 on a 2D grid of varying resolutions, the regular \ac{fmm} was evaluated for its accuracy in solving the anisotropic eikonal equation, which in our case corresponds to find \acp{at}. The errors in the solution of the eikonal equation were 40 to 50 times larger compared to those of the buffered \ac{fmm}, with the fast sweeping method, which is a classical iterative method, used as the reference solution~\cite{qian2007fast}.
 
Another constraint of the regular \ac{fmm} is the inability to simulate reentry and reactivation. Pernod et al. also developed the multifrontal \ac{fmm}, integrating a REFRACTORY list to include reentry~\cite{pernod2011multi}. However, this adaptation omitted the ``CHANGED'' list essential for managing anisotropic propagation, thereby compromising its efficacy in such scenarios. Although no reason for this omission is provided, it is possible that they faced similar challenges to those motivating this work. Specifically, changes in the solution of the eikonal equation make it difficult to determine definitive \acp{at} for calculating \acp{rt}. Gassa et al. later incorporated ideas from the multifrontal \ac{fmm}, the Dijkstra's algorithm and the Mitchell \& Schaeffer membrane model, reproducing spiral reentries in the atria~\cite{corrado2018conduction,gassa2021spiral,mitchell2003two, wallman2012comparative}. Notably, this adaptation neglected the use of recursive corrections or any other mechanism for better handling of anisotropy, possibly due to similar challenges encountered in Pernod et al.'s work. 

The numerical solution of the anisotropic eikonal equation with regular FMM-based methods without recursive corrections does not lose stability under grid refinement~\cite{sethian2003ordered}. The error from neglecting recursive corrections can be small, especially in experiments with small geometries, low anisotropy, and homogeneous properties~\cite{sethian2003ordered,cristiani2009fast} and errors may be less apparent due to the stability of the solution~\cite{sethian2003ordered}. However, arrhythmia simulations of multiple reentry cycles in realistic geometries with heterogeneous properties and high anisotropy may deviate significantly from the viscosity solution. For these complex cases, some of the adaptations made in the \ac{dream} might be particularly useful.

The \ac{fim} is better suited to solve systems with anisotropic conditions~\cite{fu2013fast}. However, the iteration process hinders the implementation of reactivation patterns and reentry. The \ac{dream} uses the \ac{cycfim} framework to manage anisotropic propagation through correction mechanisms. The \ac{dream} also addresses reentry by incorporating a ``safety margin in time'' $\tau_{\mathrm{s}}$, allowing for corrections before calculating \ac{rt} based on a reliable \ac{at}. This approach could potentially also be applied for merging the anisotropic \ac{fmm} and the multifrontal \ac{fmm}, leveraging the strengths of both methods~\cite{pernod2011multi}. However, the computational cost of buffered \ac{fmm} and the anisotropic \ac{fmm} can exceed that of iterative methods in worst-case scenarios with high anisotropy. This is due to the increase in the ``BUFFER'' and ``CHANGED'' list sizes, respectively~\cite{pernod2011multi,cristiani2009fast}. Additionally, an \ac{fim}-based algorithm is more favorable because it can be parallelized more easily while maintaining computational density~\cite{capozzoli2013comparison,jeong2008fast,fu2011fast,fu2013fast}. The \ac{dream} is the first model to use an iterative method such as \ac{cycfim}, ionic models and \ac{cv} restitution to simulate functional reentry to the best of our knowledge.


 Another method that uses the \ac{fim} in an alternative approach is the \ac{vita} method. This method employs the \ac{re} method based on the \ac{fim} approach~\cite{campos2022automated, neic2017efficient, fu2013fast}. Initially intended for studying ventricular tachycardia, \ac{vita} explores reentries using the \ac{fim}. While \ac{vita} relies on scar tissue with an isthmus to simulate anatomical reentry, the \ac{dream} is capable of simulating reentry in the absence of structural abnormalities, including functional and anatomical reentries. Additionally, the \ac{cycfim} within the \ac{dream} framework can simulate multiple \acp{at} per node per reentry, considering \ac{cv} restitution and \ac{erp} values to analyze reentry maintenance and average cycle length. Conversely, \ac{vita} does not compute multiple \acp{at} per node during reentry. Instead, \ac{vita} calculates 2 types of \acp{at} using the eikonal-based model. The first \ac{at} map identifies isochrone splitting and merging points, revealing isthmus exits where then elements are decoupled to induce unidirectional block for the second activation. This unidirectional block does not consider the electrophysiological properties at that location during repolarization. Subsequently, a second \ac{at} map is calculated for each isthmus exit, stimulating where the elements were decoupled. This second \ac{at} is the only one calculated during reentry. Then, the \ac{rtt} is defined as the time the wavefront takes to go around and reach the isthmus exit again. The \ac{rtt} serves as a surrogate marker for potential ablation targets if its value is longer than 50\,ms. One argument for using \ac{rtt} as a surrogate instead of personalized \ac{cv} restitution and \ac{erp} values is that the latter are rarely available in clinical settings. While most in silico studies do not personalize these variables but rather rely on literature references, some clinical and in silico studies have explored the effect of personalized \ac{cv} restitution and \ac{erp} values on reentry patterns~\cite{unger2021cycle,deng2017sensitivity, corrado2018work,martinez2024impact}. The \ac{dream} in contrast can incorporate personalized \ac{cv} restitution and \ac{erp} values extracted from patient measurements when available by adjusting parameters in the embedded ionic model and the COHERENCE() function.

While implementing $\tau_{\mathrm{s}}$ allows to have reentries with multiple activation cycles in iterative methods, it also entails a new challenge successfully addressed by the \ac{dream}. The \ac{cycfim} in step~$\mathrm{A_n}$ of the $n$-th \ac{dream} cycle and the \ac{rd} portion in step~$\mathrm{C}_{n-1}$ of the $(n-1)$-th \ac{dream} cycle operate within different, non-overlapping time windows. Consequently, in scenarios such as reentry, it is common for the \ac{at} of a node to be calculated in step~$\mathrm{A}_{n}$ without the previous \ac{rt} for that node having been calculated by the \ac{rd} model during step~$\mathrm{C}_{n-1}$, as required by the \ac{cycfim}. The \ac{dream} overcomes this limitation in step~$\mathrm{D}_{n-1}$ by briefly running the ionic model on the activated nodes until the voltage threshold is reached. This enables the \ac{erp} restitution to be directly obtained from the ionic model instead of relying on provided phenomenological curves~\cite{Loewe-2019-ID12386,pernod2011multi,serra2022automata}.

This study also demonstrated how the \ac{dream} could adapt to different ionic models by adjusting the corresponding parameters. This adaptability allows for the selection of the most suitable ionic model for a specific research question, depending on the required levels of computational speed and physiological detail. For instance, the Courtemanche et al. model is ideal when detailed biophysical information is needed, although it comes at the cost of computational speed~\cite{courtemanche1998ionic}. In contrast, the simplified Bueno-Orovio et al. model, which includes the main ionic currents that modify the \ac{ap} morphology and basic calcium dynamics, offers intermediate complexity with improved computational efficiency~\cite{bueno2008minimal}. At the other end of the spectrum, the simplified model of Mitchell \& Schaeffer combines fast inward and outward currents to represent basic \ac{ap} dynamics, providing higher computational speed~\cite{mitchell2003two}. 
The \ac{dream} was able to reproduce restitution behaviors very similar to the monodomain model when reducing the level of detail in the embedded ionic model for the sake of computational efficiency, as shown in the comparison between the Courtemanche et al. and Bueno-Orovio et al. ionic models. This demonstrates that the \ac{dream} is flexible enough to be used in various research scenarios where computational resources are a constraint without significantly compromising on accuracy. 

Another novel aspect of the \ac{dream} lies in its approach to incorporate \ac{cv} restitution. Previous methods typically involved incorporating the \ac{di} at every node from the previous activation cycle or the \ac{di} of neighboring nodes in the current activation cycle~\cite{Loewe-2019-ID12386,corrado2018conduction}. In this study, \ac{cv} is calculated using the \ac{di} of the current activation cycle, allowing for better adaptation to sudden changes in activation frequency and \ac{di}. Furthermore, as \ac{cv} is assigned per node rather than per element, the minimization of \ac{at} is facilitated using analytical formulas such as Eq.~\ref{eq:lat_triangle}. This eliminates the need for the Dijkstra's algorithm and additional pathways within the triangles can be considered~\cite{wallman2012comparative}. In contrast, Corrado et al. proposed incorporating the \ac{di} within the Dijkstra's algorithm~\cite{corrado2018conduction} which considers trajectories solely along the edges.

\subsection{Limitations and Future Work}\label{subsec:future}

Several attempts have been made to incorporate additional properties such as curvature and diffusion effects into the pure eikonal model~\cite{sermesant2005fast,gander2023accuracy,colli1990wavefront,keener1991eikonal}. However, existing eikonal models still struggle to accurately represent complex activation patterns, such as multi-wavelet reentry, which are influenced by high wavefront curvatures, bath loading, wave collisions and other source-sink mismatch effect. These properties are typical of propagation in fibrotic and heterogeneous tissue, a key factor in sustaining reentry~\cite{gander2023accuracy}. To simulate these complex activation patterns accurately, it is necessary to accurately model diffusion and curvature effects, as well as reactivation and repolarization.

In this work, the approximation of the diffusion current was incorporated as a function that depends solely on the \acp{at} with constant amplitude and duration. This function is unaffected by electrophysiological properties of the surrounding tissue. This limitation implies that the \ac{dream} at this moment does not consider source-sink mismatch effects in the diffusion current and in the \ac{cv}. Most of the differences observed between the \ac{dream} and monodomain model in the experiments regarding reentries can be attributed to this phenomenon. In the \ac{re}$^+$ model described by Neic et al., $ I_{\mathrm{foot}} $ (analogous to $ I_{\mathrm{diff}} $ in the \ac{dream}) is added to the parabolic portion of the \ac{rd} model. On the other hand, the \ac{re}$^-$ model replaces the diffusion term $\nabla \cdot (\boldsymbol{\sigma}_{\mathrm{i}}\nabla V_{\mathrm{m}})$ with $ I_{\mathrm{foot}} $. By doing this, the \ac{re}$^+$ model managed to obtain similar repolarization phases as those obtained in the \ac{rd} models when the tissue had heterogeneous \ac{apd}. In the \ac{dream}, $ I_{\mathrm{diff}} $ is implemented as in the \ac{re}$^-$ model, resulting in sharp repolarization gradients that preserve the \ac{apd} differences between neighboring cells. When implementing the \ac{re}$^+$ approach in the \ac{dream}, one needs to ensure that the \ac{rd} wavefront is never ahead of the eikonal wavefront to prevent major artifacts in the membrane models and their refractory behavior. This is more difficult for complex activation patterns with spatially heterogeneous curvature than for planar or radial excitation spread. For instance, during the repolarization phase, new activations (i.e., not predicted by the eikonal model) occurred when there was a strong gradient between cells with short \ac{apd} and neighboring cells with longer \ac{apd}.

While this paper is limited to triangles, the \ac{dream} algorithm can be executed in tetrahedra by employing a similar approach as described in \cite{fu2013fast} and by implementing \ac{dream}'s adaptations to the tetrahedra equation within the local solver, as performed in Eq.~\ref{eq:lat_triangle}. Nevertheless, further analysis of the computational efficiency is required for \ac{dream} simulations in volumetric meshes. While the simulations in this paper were conducted in serial code, parallelization can further enhance the efficiency of the \ac{dream}. Another potential improvement to decrease the computing time of the \ac{dream} would be to increase the integration time step when calculating the \ac{rt} in cases where they were not yet provided by the ionic model. Further improvements could be achieved in calculating the \acp{rt}. Currently, \acp{rt} are calculated twice in some nodes: once to provide the eikonal model with the necessary information and a second time when the transmembrane voltage is calculated. 

In this work, atrial cell models and \ac{cv} values were tested in the context of \ac{af}. Nonetheless, other atrial or ventricular models can also be used with the \ac{dream} as long as the parameters of \ac{cv} restitution and \ac{ap} properties are tuned accordingly. Investigating the tissue effects of additional changes in ionic models, such as pharmacological effects or channel mutations, represents another valuable area for exploration.  On the other hand, the \ac{dream} with the embedded ionic model of Mitchell \& Schaeffer demonstrated challenges in reproducing the restitution behavior at short pacing cycle lengths observed in the corresponding monodomain model. Therefore, a better adjustment of the \ac{dream} parameters is required to further improve the faithful representation of \ac{cv} restitution curve and simulation of reentrant scenarios.

More systematic analysis of the new parameters introduced in the \ac{dream} is required to further understand the optimal tuning  that allows for a good balance between accuracy and computational efficiency. We expect that $\tau_{\mathrm{s}}$ should increase when the anisotropy ratio increases as more changes are required. On the other hand, $\tau_{\mathrm{inc}}$ must be smaller than the sum of the longest possible \ac{apd} and the minimum \ac{di}, i.e, parameter $\theta$ in Eq.~\ref{eq:function_cv}. Moreover, further investigation of the COHERENCE() function could enhance the approximation of \ac{cv} restitution, particularly near propagation failure.

The \ac{cycfim} embedded in the \ac{dream} could potentially also be used as an alternative method to simulate other applications of \ac{rd} models. In this case, changing the ionic model according to the reaction part of the potential application would be necessary. Potential use cases include the \ac{rd} model simulating cyclical phenomena like the Belousov-Zhabotinsky reaction, which can also exhibit spiral propagation patterns~\cite{keener1986spiral}.

\subsection{Conclusion}\label{subsec:conclusion}
The \ac{dream} presents several advancements in simulating cardiac arrhythmias compared to existing eikonal-based models while retaining their main advantages and making it accessible as part of the openCARP simulator~\cite{openCARP-paper}. By inheriting consistency across mesh resolutions from the eikonal model, the \ac{dream} achieves faster computing times compared to \ac{rd} models for a given desired accuracy. Additionally, the \ac{dream} faithfully represents functional reentry without the need for structural abnormalities like scar tissue. The use of \ac{cycfim} enables multiple \acp{at} per node, allowing for better analysis of reentry patterns in anisotropic media. Moreover, the \ac{dream} permits modification of \ac{cv} restitution and \acp{erp}, enhancing the personalization of cardiac computer models. Ongoing work aims to overcome limitations in representing source-sink balance. Overall, the \ac{dream} offers promising prospects for advancing our understanding and improving treatments of cardiac arrhythmias compatible with clinical time frames. Finally, the \ac{cycfim} may find applications beyond cardiac modeling in simulating cyclical phenomena.

\bmhead{Acknowledgments}

This work was supported from the European Union’s Horizon research and Innovation programme under the Marie Skłodowska-Curie grant agreement No. 860974, by the Leibniz ScienceCampus ``Digital Transformation of Research'' with funds from the programme ``Strategic Networking in the Leibniz Association'', the European High-Performance Computing Joint Undertaking EuroHPC under grant agreement No 955495 (MICROCARD) co-funded by the Horizon 2020 programme of the European Union (EU) and the German Federal Ministry of Education and Research (BMBF),  by the Deutsche Forschungsgemeinschaft (DFG, German Research Foundation) – Project-ID 258734477 – SFB 1173  and LO 2093/6-1 (SPP 2311) and – Project-ID 394433254 (LU 2294/1-1, DO 637/23-1, WA 4259/1-1), and by the MCIN\textbackslash AEI and the European Union NextGenerationEU\textbackslash RTR under grant PLEC2021-007614.

\bmhead{Authors Contributions}

Conceptualisation: CBE, AL. Data curation: CBE, PMD. Formal Analysis: CBE, SA, SB. Funding Acquisition: AL. Investigation: CBE, SA, LU, SB. Methodology: CBE, JS, SA, AL. Project administration: CBE, AL. Software: CBE, JS, MH, SB, LU, SA, PMD, AL. Supervision: AL. Validation: JS, SB, SA. Visualization: CBE, SA, JK. Writing – original draft: CB, SA. Writing – review and editing: CBE, JS, MH, LU, PMD, JK, SB, AL, SA.

\bmhead{Data availability}
No original data was used for this work

\bmhead{Code availability}
The repository with all the original code is publicly available in the \href{https://git.opencarp.org/openCARP/openCARP/-/tree/openCARP_DREAM}{openCARP repository} (\url{https://git.opencarp.org/openCARP/openCARP/-/tree/openCARP_DREAM}).

\bmhead{Conflict of interest}
The authors declare no competing interests

\bibliography{sn-article}

\end{document}